%% file: Veni_Vidi_Dixi.tex
\documentclass[9pt,sigconf]{acmart}

\copyrightyear{2019}
\acmYear{2019}
\setcopyright{acmcopyright}
\acmConference[CoNEXT '19]{The 15th International Conference on emerging Networking EXperiments and Technologies}{December 9--12, 2019}{Orlando, FL, USA}
\acmBooktitle{The 15th International Conference on emerging Networking EXperiments and Technologies (CoNEXT '19), December 9--12, 2019, Orlando, FL, USA}
\acmPrice{15.00}
\acmDOI{10.1145/3359989.3365418}
\acmISBN{978-1-4503-6998-5/19/12}

\usepackage{graphicx}
\usepackage{bbm}

\usepackage{xcolor}
\usepackage{amsmath}
\usepackage{varwidth}
\usepackage{tikz}
\usepackage{pgf-umlsd}
\usetikzlibrary{shapes,arrows,shadows}

\usepackage{bm}
\usepackage{amssymb}
\usepackage{algorithm}
\usepackage{algpseudocode}
\usepackage{pgf}
\usetikzlibrary{arrows,automata}
\usepackage{pgf-umlsd}
\usetikzlibrary{calc}
\usepackage[latin1]{inputenc}
\usetikzlibrary{decorations.pathreplacing}
\usetikzlibrary{shapes.geometric}

\usepackage{subcaption}

\input{Notation}
\definecolor{TUMBlau}{RGB}{0,101,189} 
\definecolor{TUMBlauDunkel}{RGB}{0,82,147} 
\definecolor{TUMBlauHell}{RGB}{152,198,234} 
\definecolor{TUMBlauMittel}{RGB}{100,160,200} 

\definecolor{TUMElfenbein}{RGB}{218,215,203} 
\definecolor{TUMGruen}{RGB}{162,173,0} 
\definecolor{TUMOrange}{RGB}{227,114,34} 
\definecolor{TUMGrau}{gray}{0.6} 

\usepackage{amssymb}
\usepackage{pifont}
\newcommand{\cmark}{\ding{51}}%
\newcommand{\xmark}{\ding{55}}%
\usepackage{booktabs} 

\usepackage{array}
\newcolumntype{M}[1]{>{\centering\arraybackslash}m{#1}}
\newcolumntype{N}{@{}m{0pt}@{}}

\usepackage[T1]{fontenc}

\begin{document}

\title{Veni Vidi Dixi: Reliable Wireless Communication with Depth Images}

\author{Serkut Ayva\c{s}{\i}k, H. Murat G\"ursu, Wolfgang Kellerer}

\affiliation{
	\institution{
		Technical University of Munich}
}

\begin{abstract}
	
	The upcoming industrial revolution requires deployment of critical wireless sensor networks for automation and monitoring purposes. However, the reliability of the wireless communication is rendered unpredictable by mobile elements in the communication environment such as humans or mobile robots which lead to dynamically changing radio environments. Changes in the wireless channel can be monitored with frequent pilot transmission. However, that would stress the battery life of sensors. In this work a new wireless channel estimation technique, Veni Vidi Dixi, VVD, is proposed. VVD leverages the redundant information in depth images obtained from the surveillance camera(s) in the communication environment and utilizes Convolutional Neural Networks (CNNs) to map the depth images of the communication environment to complex wireless channel estimations. VVD increases the wireless communication reliability without the need for frequent pilot transmission and with no additional complexity on the receiver. The proposed method is tested by conducting measurements in an indoor environment with a single mobile human. Up to authors' best knowledge our work is the first to obtain complex wireless channel estimation from only depth images without any pilot transmission. The collected wireless trace, depth images and codes are publicly available.
\end{abstract}

\ccsdesc[500]{Networks~Network experimentation}
\ccsdesc[500]{Networks~Network reliability}
\ccsdesc[500]{Computing methodologies~Neural networks}

\keywords{Channel Estimation, Convolutional Neural Networks, Dataset}

\maketitle
 
\input{intro}

\input{background}
\input{meas}

\input{model}
\input{chan_chan}
\input{eval}
\input{sota}
\input{Conc}
\section*{Acknowledgment}
The authors would like to thank Dr.-Ing. Christoph Bachhuber for his valuable support with the visual data collection. This work has received funding by the Bavarian Ministry of Economic Affairs, Regional Development and Energy as part of the project `5G Testbed Bayern mit Schwerpunktanwendung eHealth'.

\input{app}

\bibliographystyle{acm}
\bibliography{Veni_Vidi_Dixi}

\end{document}

%% file: intro.tex
\section{Introduction}
\label{sec:intro}

Wireless communication is one of the key enablers for industrial revolution. However, the current capabilities of wireless communication is not sufficient to support the reliability requirements of critical communications imposed by automation systems. The main reason for the lack of reliability is the uncontrolled nature of the wireless channel. 

The state of the art framework to guarantee reliability of wireless communication requires transmission of known signals over the wireless channel, called channel sounding, for obtaining fresh information about the wireless channel, called channel estimation. Known signals, pilots, are transmitted frequently to update this estimation. Following, the channel capacity is calculated and an appropriate coding and modulation scheme is selected to guarantee the reliability of the communication. Finally, the packet is transmitted, received and decoded reliably. 

However, there are many problems with this traditional approach. First, it assumes that the channel does not vary from the transmission of the pilot until the reception of the packet, which puts a strict constraint on timing \cite{rappaport1996wireless}. Second, this multi-step framework can be too slow for certain applications such that the data needs to be transmitted before the hand-shake; hence the reliability cannot be guaranteed \cite{pocovi2018achieving}. Third, a separate pilot has to be sent for each connection. Obviously in future, with the vast amount of sensors connected to a single station such an approach does not scale \cite{gursu2018multiplicity}.
\begin{table}[t!]
	\centering
	\vspace{0.5cm}
	\begin{tabular}{||c | c c c||} 
		\hline
		Technique & Reliable & Scalable & Dynamic \\ [0.5ex] 
		\hline\hline
		Blind & \xmark & \cmark & \cmark \\ 
		\hline
		Pilot & \cmark & \xmark & \cmark \\
		\hline
		Time-Series & \cmark & - & \xmark \\
		\hline
		VVD & \cmark & \cmark & \cmark \\ \hline
	\end{tabular}
	\caption{Comparison of channel estimation techniques}
	\label{tab:summary}
\end{table}
\raggedbottom

\begin{figure*}[t!]
	\hspace{-0.5cm}
	\begin{subfigure}[t]{0.32\textwidth}
		\includegraphics[scale=0.32]{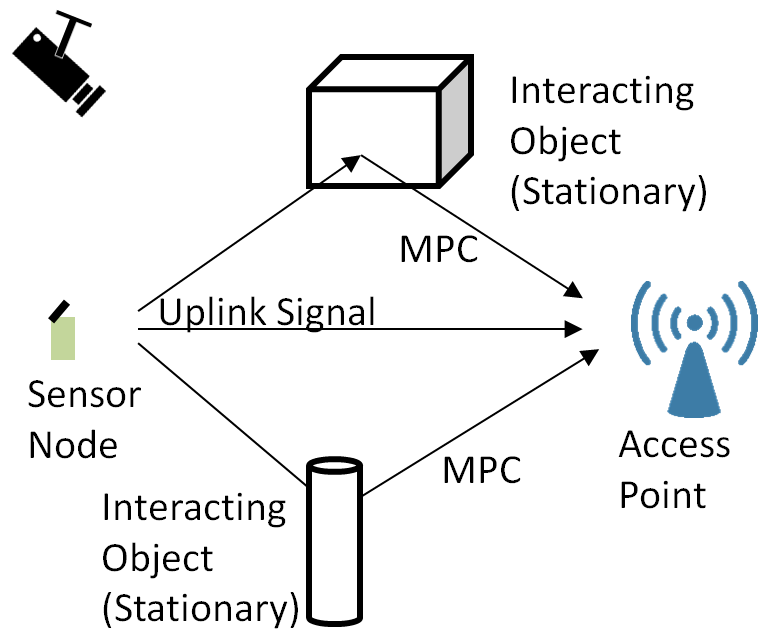}
		\centering
		\caption{Stationary environment}
		\label{fig:SIC_1}
	\end{subfigure}
	\hspace{0.2cm}
	\begin{subfigure}[t]{0.32\textwidth}
		\includegraphics[scale=0.32]{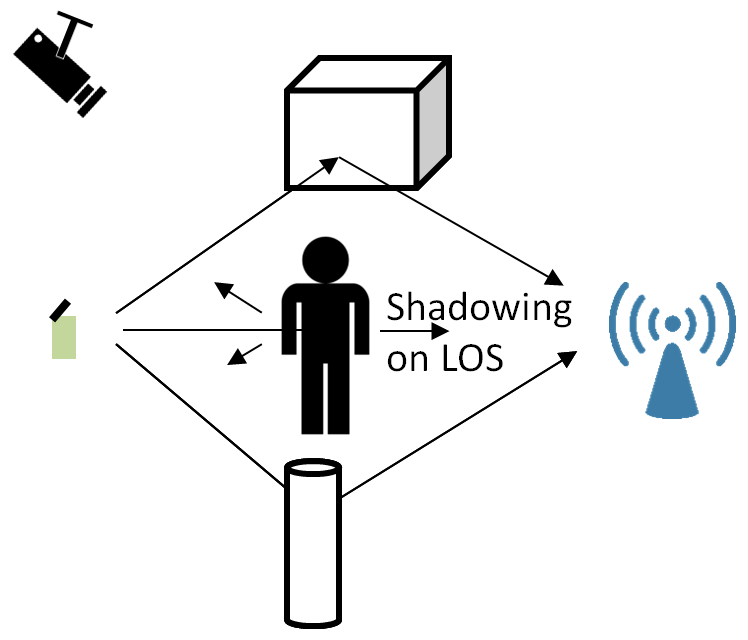}
		\centering
		\caption{Human interaction distorts the LoS signal}
		\label{fig:SIC_2}
	\end{subfigure}
	\hspace{0.2cm}
	\begin{subfigure}[t]{0.32\textwidth}
		\includegraphics[scale=0.32]{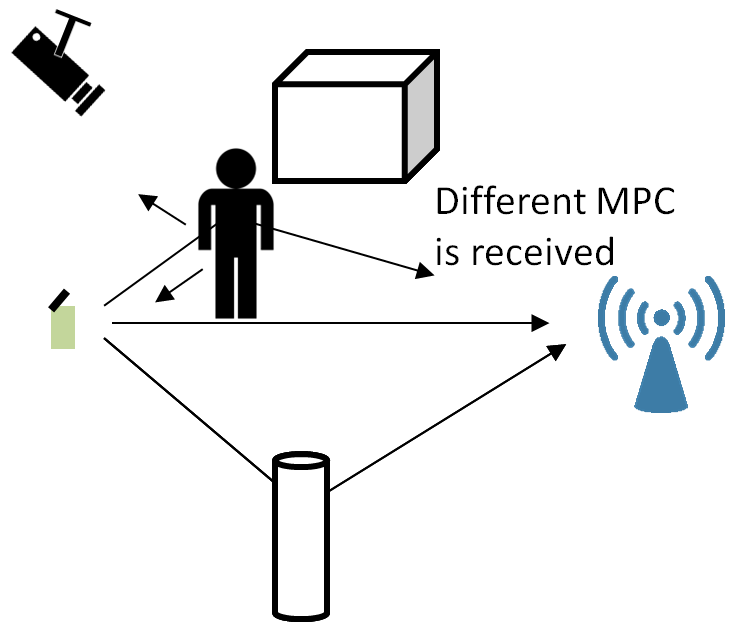}
		\centering
		\caption{Human interaction distorts an MPC}
		\label{fig:SIC_3}
	\end{subfigure}
	\caption{Variations of MPCs with respect to interacting object under camera inspection}
	\label{fig:Motivation}
\end{figure*}

A solution to these problems is to decode with imperfect channel estimation \cite{do2016uplink}. The performance of this approach depends on the dynamics of the physical environment where the wireless communication takes place. Hence, it cannot guarantee reliable communication. Another approach, as in \cite{han2004channel}, is to use time-series estimation to model the behavior of the wireless channel. Via decreasing the burden on pilot transmission this approach provides a trade-off between blind and estimation based decoding. This approach can adjust the frequency of the control information but this should depend on the mobility in the environment. Fixing the pilot transmission to a certain periodicity suffers from dynamic environment. The discussed disadvantages of the channel estimation techniques are summarized in Tab.~\ref{tab:summary}. 

In this work, we propose a new solution to this problem, Veni Vidi Dixi (VVD), I came, I saw and I said. VVD uses depth images captured from the communication environment to estimate the wireless channel. This enables reliable decoding of the packets without the need for any pilot overhead. 
Our contributions can be summarized as below:

\begin{enumerate}
\item We introduce a novel algorithm, VVD, that performs blind channel estimation in terms of complex channel coefficients from depth images of the communication environment.
\item We collect and publish a wireless trace together with synchronized camera depth images open for further research.   
\item We compare VVD with other blind time series estimation techniques such as Kalman filter and show that on average it performs better without the need for pilot update.
\end{enumerate}

In Sec.~\ref{sec:motivation} background related to hypotheses of the work is introduced. In Sec.~\ref{sec:measurement} the measurement set-up is described and hypotheses are tested. In Sec.~\ref{sec:algorithm} VVD algorithm is introduced. In Sec.~\ref{sec:comparison} baseline algorithms are explained together with the comparison metrics. Algorithms are compared and evaluated in Sec.~\ref{sec:eval}. Discussions stemming from the results are introduced in Sec.~\ref{sec:discs}. Related work is summarized in Sec.~\ref{sec:sota} and the paper is concluded with a summary in Sec.~\ref{sec:conc}.

Our wireless trace, depth images, codes as well as Machine Learning (ML) model and their detailed usage explanation are publicly available at: {https://gitlab.lrz.de/lkn\_measurements/vvd\_measurements}.

%% file: background.tex
\section{Background and Hypotheses}
\label{sec:motivation}

The wireless channel is roughly characterized by large-scale fading and small-scale fading \cite{Tse04}. The large-scale fading is affected by the distance between transmitter and receiver and by objects shadowing the channel. Whereas small-scale fading originates from different paths between transmitter and receiver due to reflection and diffraction of the transmitted signal. Each replica of the signal that travel a different path is called a Multi Path Component (MPC). At the receiver MPCs interfere with each other at different times with different amplitudes and phase shifts. The MPCs in a wireless communication setting is illustrated in Fig.~\ref{fig:Motivation}. In Fig.~\ref{fig:SIC_1} a scenario with 3 distinct MPCs, 1 line of sight (LoS) and 2 Non-LoS (NLoS), is illustrated. In Fig.~\ref{fig:SIC_2} a human is blocking the LoS and in Fig.~\ref{fig:SIC_3} a human is blocking a NLoS MPC.
 
Hence, in order to correctly interpret the received signal, receiver needs to estimate these effects, i.e., perform wireless channel estimation. 

\subsection{Channel Estimation}

The wireless channel can be modeled by an impulse response and therefore can be assumed as a linear filter  \cite[p.~10]{Tse04}. Considering a linear filter interpretation of the wireless channel, the relation between transmitted and received signal can be simply explained by Equation \ref{eq:channel}  \cite[p.~21]{Tse04} for linear time-variant (LTV) systems:

\begin{equation}
y(t) = \int_{-\infty}^{\infty} x(t-\tau)h(t,\tau)d\tau,
\label{eq:channel}
\end{equation}

where ${y(t)}$ denotes the received signal at time instant $t$, ${x(t)}$ corresponds to the transmitted signal, ${h(t,\tau)}$ represents channel impulse response (CIR) at time instant $t$. Assuming limited mobility, we use a quasi-static, block fading channel model such that ${h(t,\tau)}$ does not vary with $t$ for a block. It is clear that to obtain ${x(t)}$ from ${y(t)}$, ${h(t,\tau)}$ is needed. 

Before delving into the extraction of the channel information from a received signal, it is vital to comprehend the basic relation of ${h(t,\tau)}$ to the MPCs. In Equation \ref{eq:channel2}, a impulse response representation of fading multipath channel, ${h(t,\tau)}$, can be observed  \cite[p.~21]{Tse04}:

\begin{equation}
h(t,\tau) = \sum_{i}^{N} a_{i}(t)\delta(\tau-\tau_{i}(t)), \footnote{This representation is also known as tapped delay line.}
\label{eq:channel2}
\end{equation}

${a_{i}(t)}$ are the complex coefficients for the taps where each tap represents the sum of several closely spaced MPCs, ${\tau_{i}(t)}$ corresponds to the delay of ${i^{th}}$ tap, ${N}$ denotes the total number of taps. 

There are multiple ways to obtain ${h(t,\tau)}$. In this work, we are focused on blind channel estimation, and pilot-aided channel estimation which is also known as training-based channel estimation. Training based channel estimations are based on the idea of an agreement between the transmitter and the receiver such that a portion of the received signal is already known at the receiver. 

It is worth mentioning that for practical processing of the wireless signal, a discrete time model is considered as given with Equation \ref{eq:channel_discrete} where $h_{l}[m]$ is the $l^{th}$ tap of the Finite Impulse Response (FIR) filter at sample index $m$ \cite[p.~26]{Tse04}:

\begin{equation}
y[m] = \sum_{l}^{N}h_{l}[m]x[m-l].
\label{eq:channel_discrete}
\end{equation}

Assuming block fading channel, the wireless channel, $h_{l}[m]$, becomes $h_{l}^{k}$  for $s$ samples of the block $k$, i.e, $ m~\in~\{1, 2,... , s\}$. In other words, the wireless channel is time-invariant for one received block (packet frame) but time-variant with respect to different packets, denoted with $()^k$. The linear least-squares (LS) estimate is used to obtain the FIR filter throughout the work. With the quasi-static channel assumption, for one received block LS estimate can be written as:

\begin{equation}
\hat{\boldsymbol{h}}_{LS}^{k} = ({(X^{k})^H}X^{k})^{-1}{(X^{k})^H}\boldsymbol{y^k},
\label{eq:LS_estimate}
\end{equation}

where $()^H$ denotes Hermitian transpose, $X^{k}$ is the convolution matrix of transmitted signal $x^{k}$ for a block duration, i.e. pilot samples. $X^{k}$ is a $(N+M-1)\times N$ matrix as shown in Equation \ref{eq:X_matrix}:

\begin{gather}
X^{k} = \begin{bmatrix}
x^{k}[1] & 0 & 0 \\
x^{k}[2] & x^{k}[1] & 0 \\
x^{k}[3] & x^{k}[2] & x^{k}[1] \\
... & ... & ... \\
x^{k}[M] & x^{k}[M-1] & x^{k}[M-2] \\
0 & x^{k}[M] & x^{k}[M-1] \\
0 & 0 & x^{k}[M]
\end{bmatrix},
\label{eq:X_matrix}
\end{gather}
where $M$ defines the number of pilot samples, $x[i]$ denotes the $i^{th}$ reference sample. $\boldsymbol{y^{k}}$ is the received signal for the duration of pilot samples, i.e. $\boldsymbol{y^{k}} = [y^{k}[1],y^{k}[2],...,y^{k}[M+N-1]]^T$. $N$ defines the total number of filter taps for the desired channel estimation. For instance, the Equation \ref{eq:channel} would estimate a 3-tapped FIR filter for an $X^{k}$ as given in Eq.~\ref{eq:X_matrix}. The selection of $N$ depends on the excess delay of the channel and the sampling rate at the receiver.

Equalization is the inversion of the estimated wireless channel which would result in a relation as in Equation \ref{eq:eq1}:  
\begin{equation}
H^k\boldsymbol{c}^k = \textbf{u}, 
\label{eq:eq1}
\end{equation}
where \textbf{u} denotes the vector that contains all zeros but only one `1' valued tap, i.e. $\boldsymbol{u} = [0,0,...,1,...,0,0]$, that is used to define the number of pre-cursor and post-cursor taps of the equalizer, $H^k$ is the convolution matrix of the estimated FIR filter with the size $(L+N-1)\times L$ where $L$ corresponds to the total tap length of $\boldsymbol{c}^k$ which is not necessarily the same as the previously introduced variable $N$. $\boldsymbol{c}^k$ is the FIR filter interpretation of channel equalizer that ideally inverts the channel. However, the estimation of the wireless channel is obtained as linear FIR filter which might not have a direct inversion. In order to find $\boldsymbol{c}^k$ again LS estimate can be used as in Equation \ref{eq:LS_estimate} as follows:
\begin{equation}
\hat{\boldsymbol{c}}_{LS}^k = ({(H^{k})^H}H^{k})^{-1}{(H^{k})^H}\textbf{u}.
\label{eq:LS_estimate2}
\end{equation}

$H^k$ is built in a similar fashion to Equation \ref{eq:X_matrix}, with tap values of $h_l^k$. In other words, instead of $x[i]$, $h_{[i]}^k$ is used. This approach is known as LS based Zero-Forcing (ZF) channel equalization. Although ZF approach is known as a noise amplifying technique in a strong fading channel, it is optimum for intersymbol interference elimination \cite{Molisch2011}. For the rest of the work, after getting the channel estimates, ZF approach is used for the recovery of transmitted signals from received signals.

\subsection{Hypotheses}

In this section we explain how we build up our proposal from the given multipath model in wireless communications. 

We build on the thesis:  `A static indoor environment (no mobility, no displacement, time change) results in static MPCs.' as evaluated in detail in \cite{puccinelli2006multipath}. Intuitively, we extend this thesis to:
\begin{enumerate}
	\item Having mobility in an indoor environment where some objects change places,  (mobility, displacement, time change) changes the phase and the amplitude of MPCs.
	\item Having mobility in an indoor environment where between two different time instances no objects change places, (mobility, no displacement, time change) the phase and the amplitude of all MPCs in both instances are similar. 
\end{enumerate}

If these hypotheses hold we can confidently say that camera images, that capture all the mobility in an indoor environment can be used to infer channel impulse response of MPCs. 
In the next section we test our hypotheses to justify using camera images.

%% file: meas.tex
\section{Measurement}
\label{sec:measurement}

\begin{figure}[t!]
	\centering
	\includegraphics[width=0.47\textwidth]{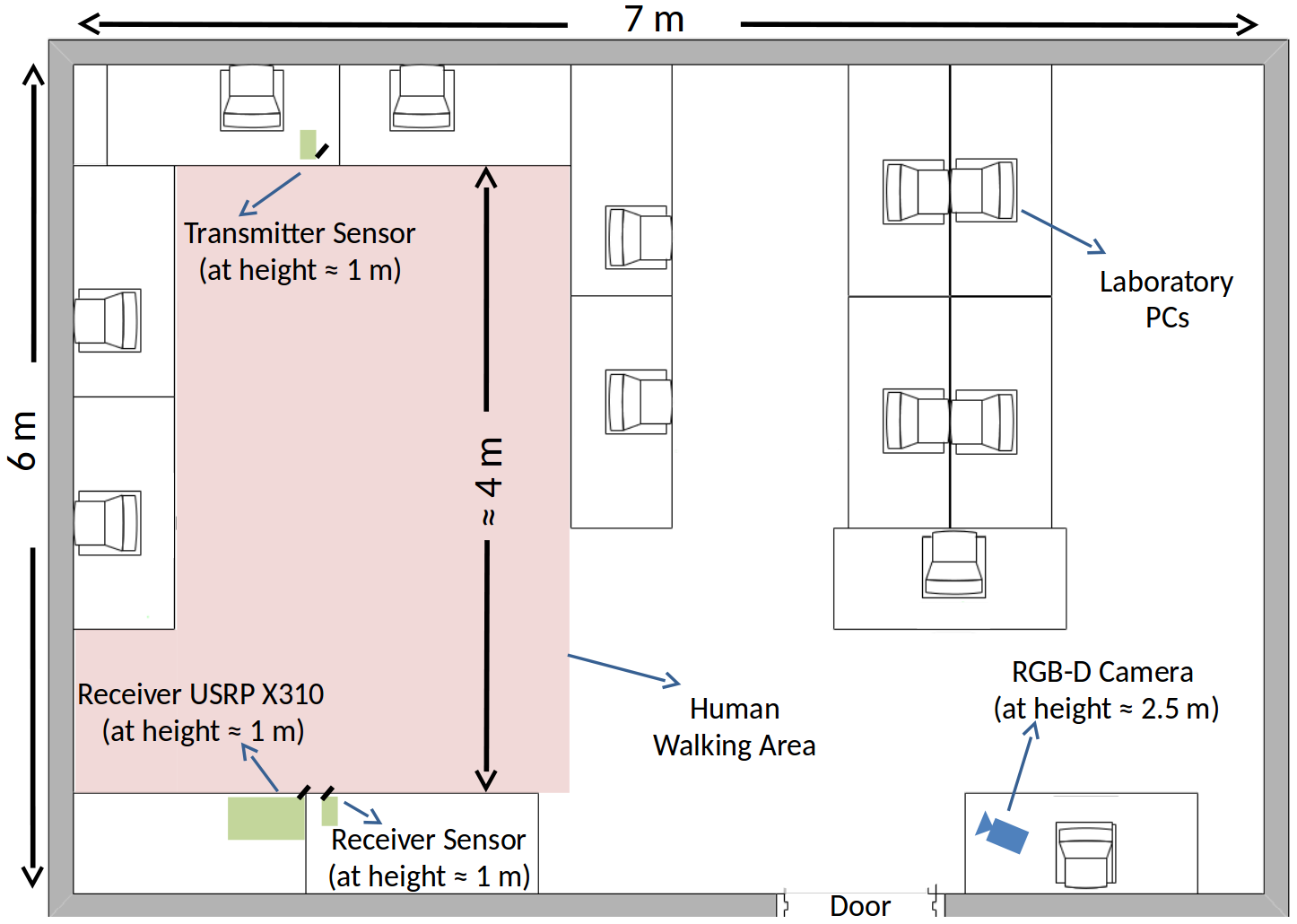}
	\caption{Overview of the Measurement Setup}
	\label{fig:Hardware_Lab}
\end{figure}

The measurements are conducted in a laboratory room (indoor environment) where there exist several PCs and metallic objects (such as robots similar to industrial environment). In Fig.~\ref{fig:Hardware_Lab} an overview of the room indicating the locations of the deployed transmitter, receiver, and the camera is displayed. Note that the sizes of the PCs, sensors, USRP, and the RGB-D camera are not scaled in this sketch. All objects are immobile and static except a single human. The movement area of the human is limited as specified in Fig.~\ref{fig:Hardware_Lab} to guarantee that all the movements are captured by the camera. The human is always mobile during the measurements.

The transmitter and receiver both use the IEEE 802.15.4 standard. The modulation scheme used by the standard is Offset-Quadrature Phase Shift Keying (O-QPSK) PHY layer at 2.45 GHz based on Direct Sequence Spread Spectrum (DSSS) technique.  DSSS is a spread spectrum modulation which spreads the information signal with a high chip rate spreading Pseudo-Noise(PN) code. Chip here denotes a sample period of spread spectrum which is smaller than the sample (bit) period of information signal. In this standard, four bits are grouped to form a symbol and then each symbol is mapped to a PN Sequence of 32 chip length. There exists an orthogonal set of 16 different PN Sequences for 16 symbols \cite{IEEE_2003}. Moreover, channel 26 of 2.4 GHz band was used for transmission as there exists 8 MHz separation between the nearest 802.11 channel in order to have a reliable operation even in the presence of WLAN interference \cite{Petrova2006}.

\begin{figure}[t!]
	\centering
	\includegraphics[width=0.47\textwidth]{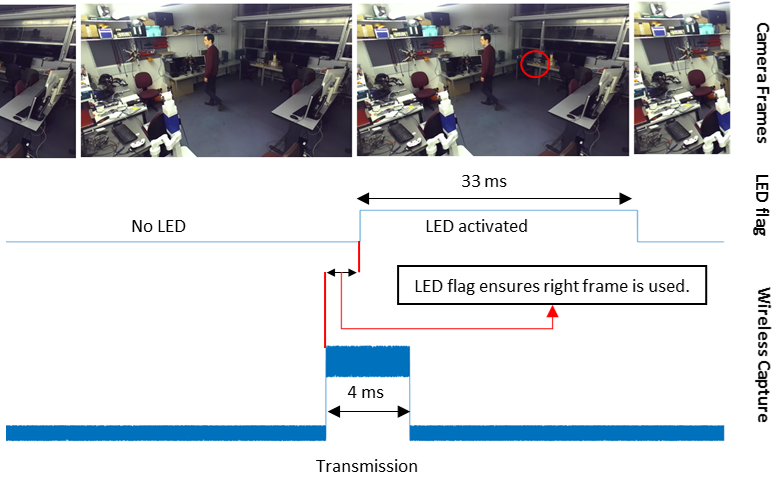}
	\caption{Wireless signal and camera synchronization}
	\label{fig:led_flag}
\end{figure}

Each 100 ms a packet is sent from the transmitter to the receiver. All of the transmitted packets are 127 Bytes in length (PSDU length) and have the same payload except the sequence number and the cyclic redundancy check. Both receiver and transmitter are Zolertia RE-Mote devices. Furthermore, both receiver and transmitter sensors are operated on battery power to reflect real deployment scenarios. Nevertheless, in order to provide an extra check for the timing of the transmissions, both the transmitter and receiver motes blink their LEDs during transmission and reception, respectively. The blink of the LEDs can be caught by the camera to synchronize one frame to the related packet. The use of LED flag catching the LED blink in the camera frames is illustrated in Fig.~\ref{fig:led_flag}. Due to the frame rate we have a frame each 33 ms. Thus, two frames can be a candidate for the same signal as illustrated in Fig.~\ref{fig:led_flag}. Use of the LED blink solves the synchronization scenario easily to guarantee that the right frame is matched to wireless transmission. 

For all measurements a USRP X310 with CBX-120 daughterboard is used as a sniffer to collect raw measurement data and to send the data to PC without any overflow. Sampling rate is 10 MHz and downsampled to 8 MHz in GNU Radio. Downsampling operation is performed due to observed frequent overflows with sampling rates above 8 MHz.\footnote{The sampling rate kept at 10 MHz on the device since having directly 8 MHz sampling with USRP X310 would result in Cascaded Integrator-Comb (CIC) filter roll-off in the passband \cite{EttusX310}} Although the channel bandwidth of IEEE 802.15.4 is only 2 MHz, more baseband bandwidth is used to increase the temporal resolution to ensure more information about the wireless channel is obtained. Higher temporal resolution allows better channel impulse response estimations by having higher multipath resolution, especially for our approach in estimating the multipath channel as an FIR filter \cite{Firooz2010}. 

The RGB-D camera used in the measurements is Stereolabs ZED camera which was operated at 30 fps at 720p video mode resembling a surveillance camera.

In the provided dataset there exist overall 22,704 packets' raw signal samples, corresponding LS estimations implemented as FIR filters at 11-Tap length, and images of each recorded measurement video. We project that depending on the environment such as large scale setups or richer multipath effects, the channel estimations will vary in a wider range. Hence, a larger dataset size would be required to learn a wider set of possible realizations.

\begin{figure*}[t!]
		\centering
	\begin{subfigure}[t]{0.3\textwidth}
		\includegraphics[width=1\textwidth]{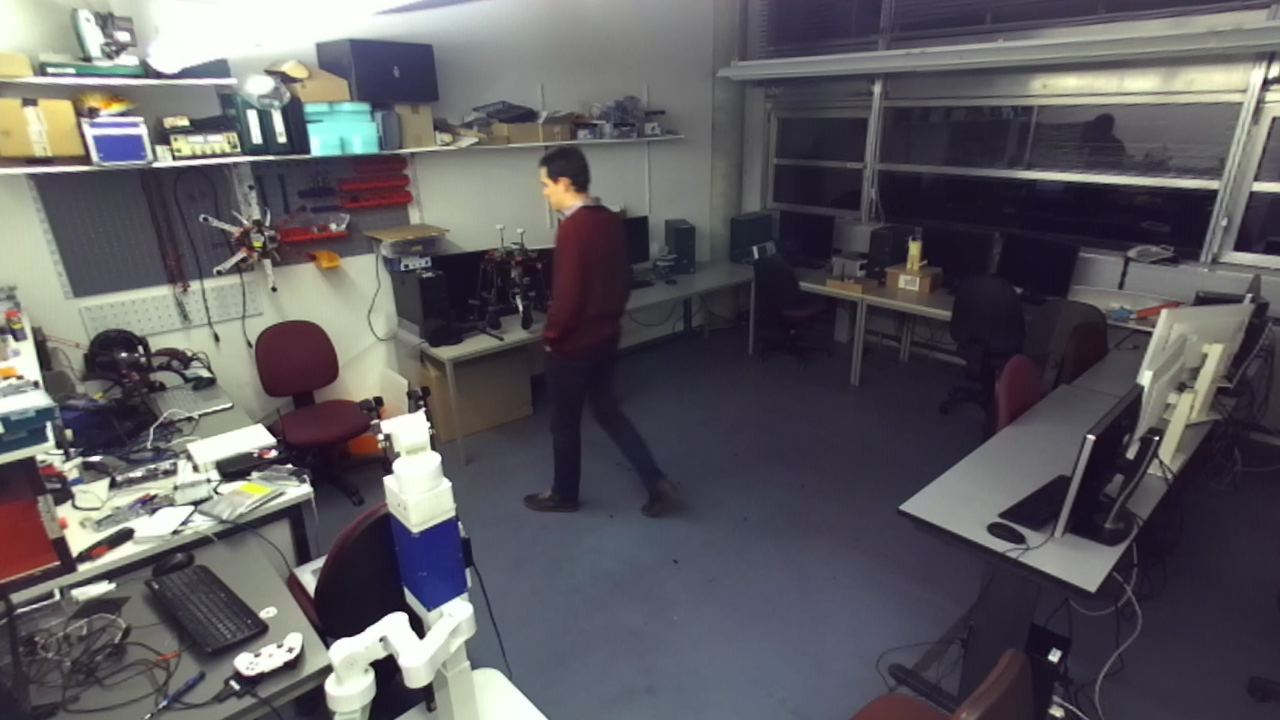}
		\caption{Control \\ Frame 497 from Set2 }
		\label{fig:Meas_set2_im}
	\end{subfigure}
	\centering
	\begin{subfigure}[t]{0.3\textwidth}
		\includegraphics[width=1\textwidth]{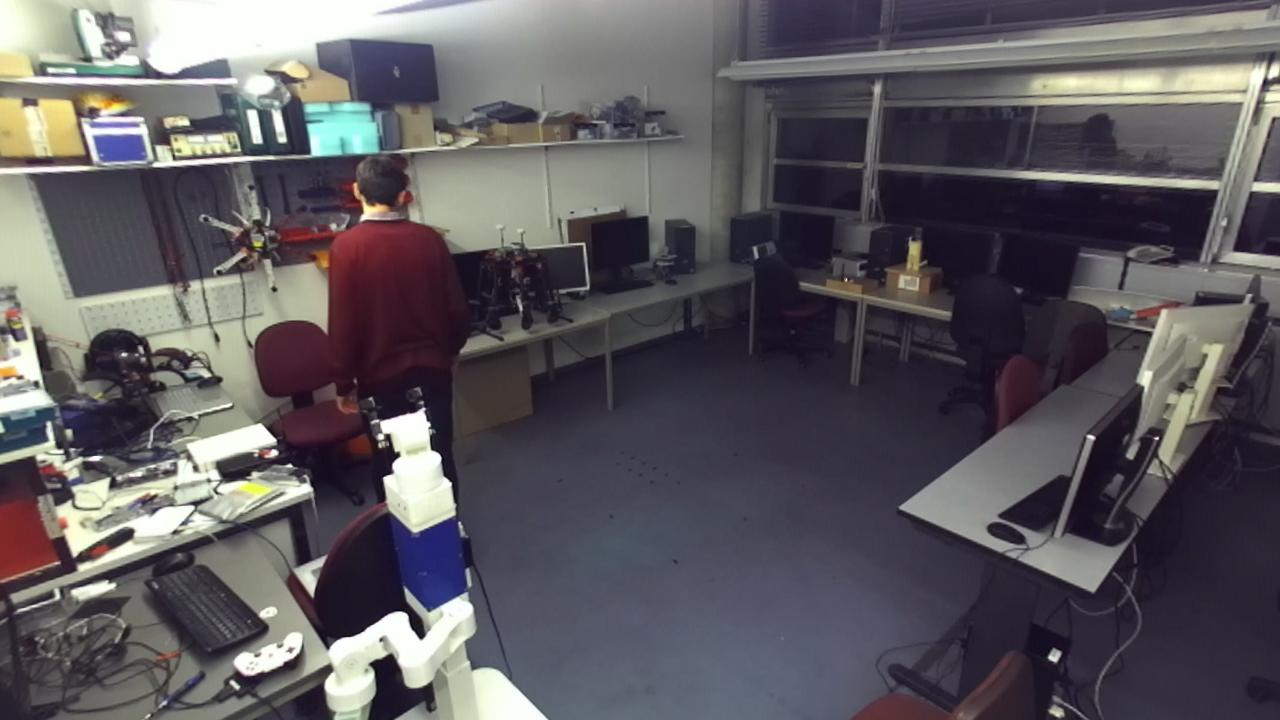}
		\caption{Hypothesis 1\\ Frame 780 from Set5}
		\label{fig:Meas_set5_im}
	\end{subfigure}
	\centering
	\begin{subfigure}[t]{0.3\textwidth}
		\includegraphics[width=1\textwidth]{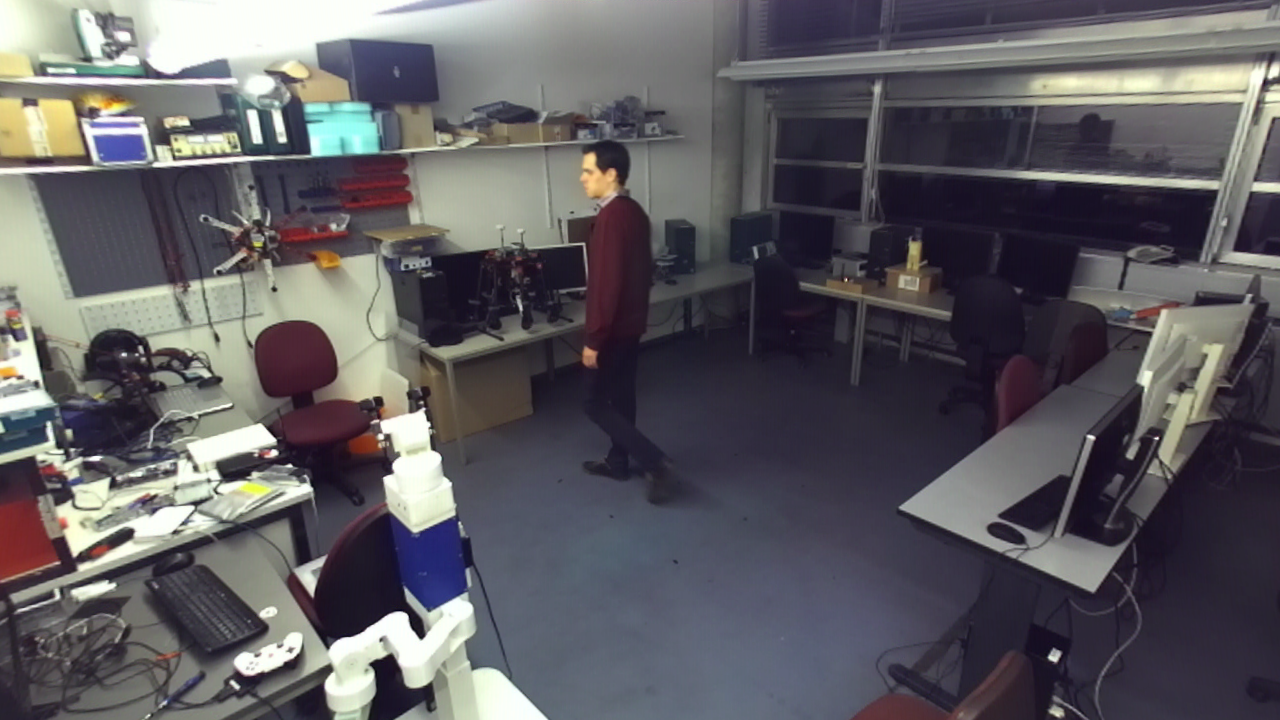}
		\caption{Hypothesis 2\\ Frame 4266 from Set5}
		\label{fig:Meas_set5_im2}
	\end{subfigure}
	\caption{Camera images for hypothesis testing. Two similar and one unrelated displacement.}
	\label{fig:Same_environment_meas}
\end{figure*}

\begin{figure*}[t!]
	\hspace{-1.8cm}
	\begin{subfigure}[t]{0.49\textwidth}
		\includegraphics[width=1.18\linewidth]{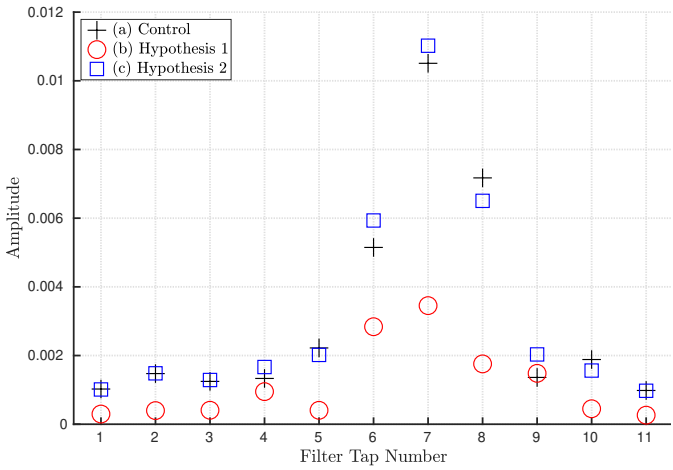}
		\centering
		\caption{Tap Coefficients in Absolute Value}
		\label{fig:Abs_taps}
	\end{subfigure}
	\hspace{0.4cm}
	\begin{subfigure}[t]{0.49\textwidth}
		\includegraphics[width=1.18\linewidth]{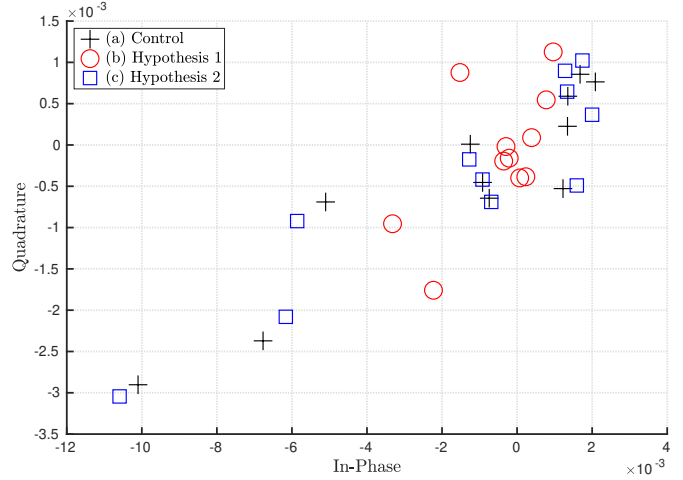}
		\centering
		\caption{Constellation Diagram of Tap Coefficients}
		\label{fig:Constellation_taps}
	\end{subfigure}
	\caption{Complex Channel Tap Coefficients in Absolute Value and Constellation Diagram for the two similar environments}
	\label{fig:Comparison_Channel}
\end{figure*} 

\subsection{Hypothesis Testing}

In this section we do preliminary tests related to our hypotheses. As all hypotheses are based on the mobility and the wireless channel state, we use camera captured images to check the state of the mobility and we use channel impulse responses captured by USRP. Between different experiments the mobility of the human changed to collect evidence related to the hypotheses.

\textit{\textbf{Hypothesis 1:} Mobility changes the phase and the amplitude of MPCs.} \\

\textit{\textbf{Hypothesis 2:} If mobile objects end up in the same place in two different time instances the phase and the amplitude of all MPCs in both instances are similar.}
\\ 

In this setting we collected measurements with random movements of the human.

In Fig.~\ref{fig:Same_environment_meas} three environments of different measurement takes are displayed as colored images. Specifically, Fig.~\ref{fig:Meas_set2_im} is from "Set2" and is used as a control case,  Fig.~\ref{fig:Meas_set5_im} and Fig.~\ref{fig:Meas_set5_im2} are from "Set5" that is recorded with an hour of difference. The former represents a test for \textit{hypothesis 1} which includes a displacement and mobility. The latter represents the \textit{hypothesis 2} where the displacement is similar to the control image but there has been mobility in between. In the Fig.~\ref{fig:Meas_set2_im} and \ref{fig:Meas_set5_im2} the human is almost equidistant to both transmitter and receiver blocking the LoS from a distance. In the Fig.~\ref{fig:Meas_set5_im} the human is standing in front of the receiver affecting the LoS and multiple MPCs. 

In Fig.~\ref{fig:Comparison_Channel} the channel estimations of all instances are given. In Fig.~\ref{fig:Abs_taps} each tap is given with its absolute distance and in Fig.~\ref{fig:Constellation_taps} each tap is displayed in terms of quadrature and in-phase with a scatter plot.

In Fig.~\ref{fig:Abs_taps} the x-axis depicts $11$ different taps, while the y-axis depicts the amplitude of each tap. First evaluating the hypothesis 1, we see that for the dominant paths, that are taps 6, 7 and 8 \footnote{This is due to our estimation in which we allow pre-cursor taps.}, the amplitude shows a difference between the control frame (a) and (b). The non-dominant taps looks close. For the hypothesis 2, we see that the tap amplitudes are a lot closer but there is no perfect match. To represent the two dimensional validation of the channel estimates, in-phase and quadrature, it is more valuable to investigate Fig.~\ref{fig:Constellation_taps}. In Fig.~\ref{fig:Constellation_taps} the validation of both hypotheses is demonstrated clearly.

The results for constellation are displayed after the mean phase shift is corrected for each estimate. The phase shift is due to imperfect crystals of the sensors \cite{1090326}. Yet, it is observed that the existing phase shift for Hypothesis 2 is only a mean phase shift on all of the filter taps such that when it is reverted on all of the taps on one estimation the two estimations become similar in constellation.

The mean phase shift can be obtained with the proposed method in \cite{CRDSA} for the respective channel estimation by using correlation. The method can be written as in Equation~\ref{eq:CRDSA} \cite{CRDSA} for two channel estimations of different instances:
\begin{equation}
\hat{\theta} = arg\{\boldsymbol{\hat{h}}_{LS}^{k_1} \cdot \{\boldsymbol{\hat{h}}_{LS}^{k_2}\}^H\},
\label{eq:CRDSA}
\end{equation}
$\boldsymbol{\hat{h}}^{k_1}$ corresponds to the estimation for the current instance and $\boldsymbol{\hat{h}}^{k_2}$ corresponds to the estimation for the older instance, and $\hat{\theta}$ denotes the phase difference between two estimations.\footnote{Same method is used to correct the mean phase shift between the obtained channel estimation from VVD and the received block by exploiting the known parts of the signal such as preamble.}

As we have approved all our hypotheses, we can test if a learning algorithm can learn the channel behavior related to camera images.

%% file: model.tex
\section{Veni Vidi Dixi Algorithm}
\label{sec:algorithm}
 
\begin{figure}[t!]
	\centering
	\includegraphics[width=0.47\textwidth]{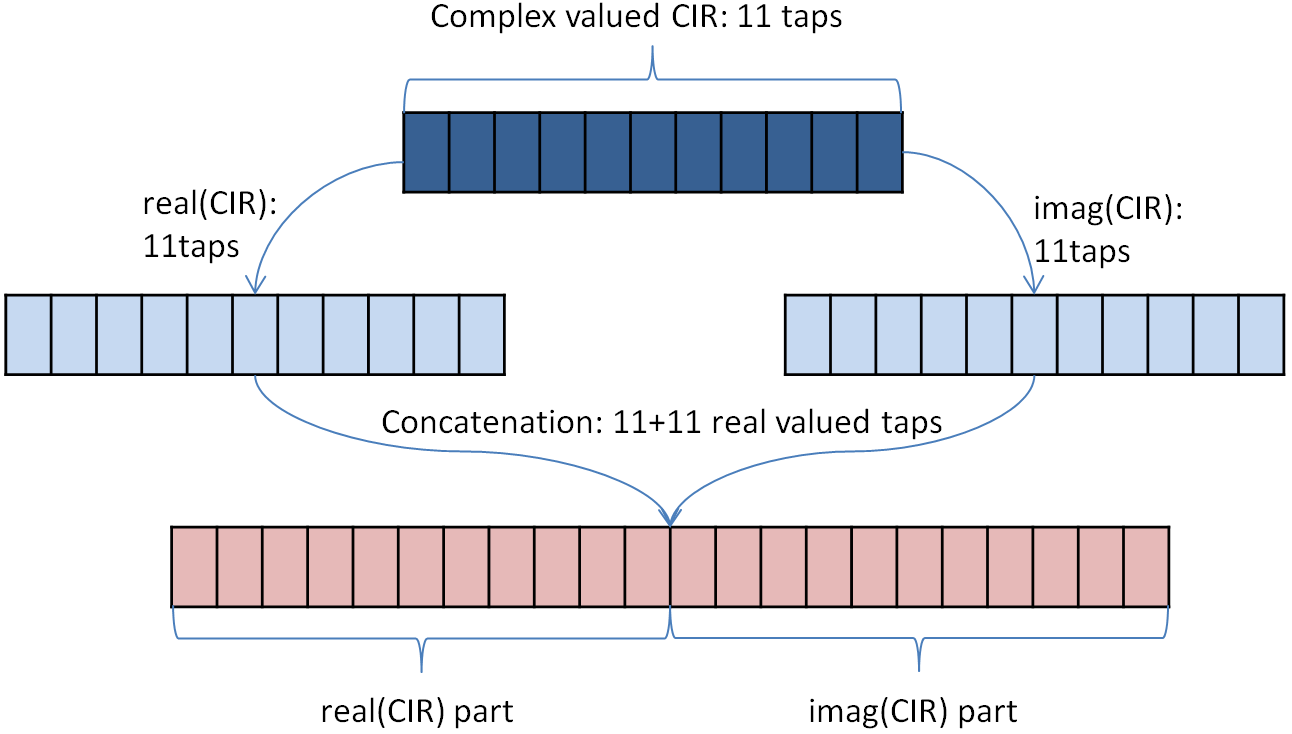}
	\caption{Complex valued CIR to real valued CIR}
	\label{fig:ML_tap_concatenation}
\end{figure}

\begin{figure*}[t]
	\centering
	\includegraphics[width=0.9\textwidth]{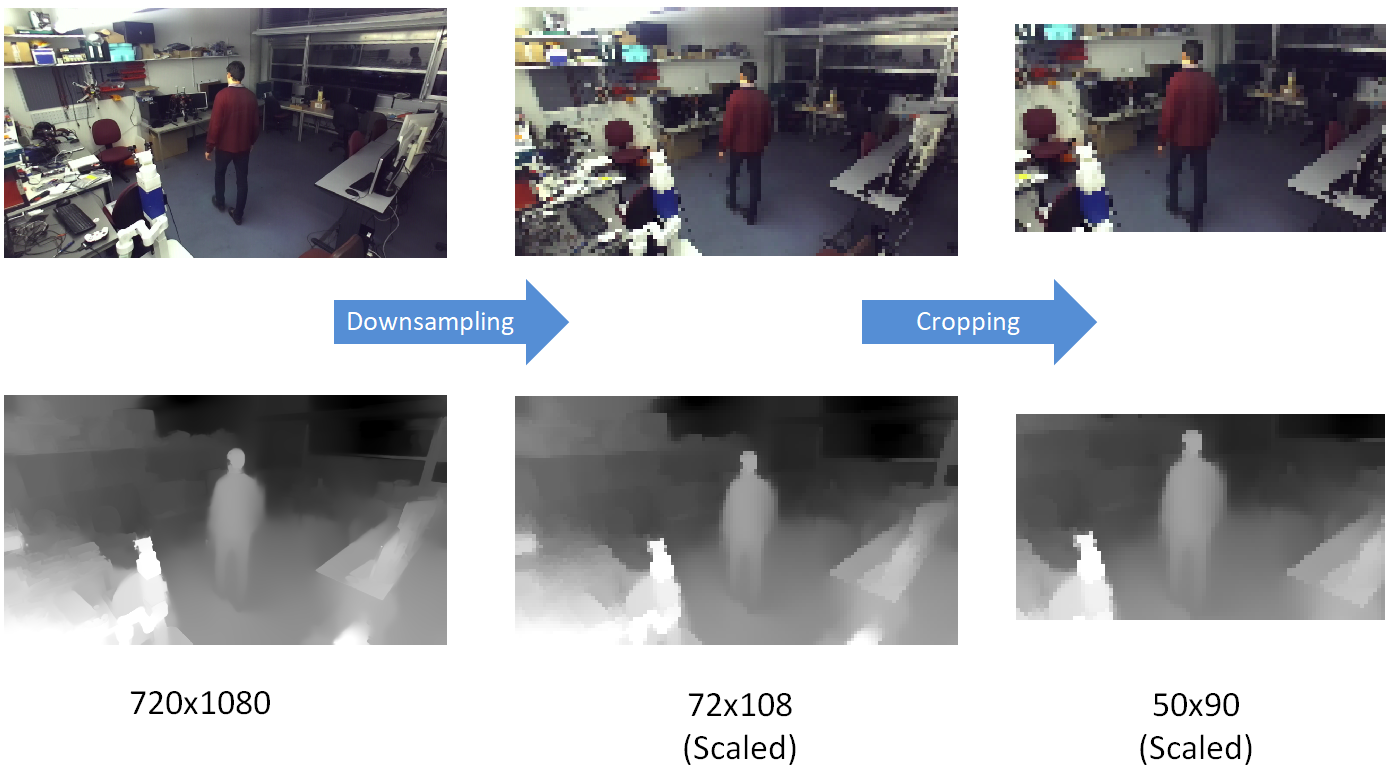}
	\caption{Image pre-processing composed of downsampling by 10 and cropping}
	\label{fig:Image_preprocessing}
\end{figure*}

The use of complex values in convolutional neural networks is still an active research topic and requires changes in the network layers \cite{Trabelsi2017}. On the other hand, wireless channel estimation has complex valued taps. We have separated the real and imaginary parts of the complex valued taps to solve the issue of complex signals. In other words, the real part and imaginary part of the complex valued taps are concatenated and all taps are treated as real valued taps as illustrated in Fig.~\ref{fig:ML_tap_concatenation}. For example, for the algorithm that is designed to learn a CIR of 11 taps, the number of neurons in the last layer is selected as 22.

The inputs of the Machine Learning (ML) algorithm are the depth images of the environment. The use of the depth images is to gain positional information of the interacting objects in the environment. The resolution of the collected images during a measurement are 720x1080, the size of all the images were reduced/downsampled to 72x108 to reduce the computational complexity \footnote{Using large image size in a ML model fills the GPU memory which forces to use small batch sizes during training. This would result in quite long training times and large ML network as it would have more parameters than a model using smaller image.}. Further the image is cropped to exclude the unrelated parts in the images \footnote{Unrelated here means that there never exist any mobile objects that would result in variations in the wireless channel. Thus, the cropped parts of the images would simply provide no extra information.} resulting in an input size of 50x90. The applied downsampling and cropping are visualized in Fig.~\ref{fig:Image_preprocessing} for one image of the measurements. In Fig.~\ref{fig:Image_preprocessing} top part displays the RGB image, whereas bottom part demonstrates the depth image for that instant.

CNN is used to extract the mobility information from the depth images. Convolutional layers in this neural network compares a pixel with its neighbors and tries to extract higher layer information through repetition of these convolutional layers. Rectified Linear Units (ReLUs) are used for activation. 

\begin{figure*}[t!]
	\centering
	\includegraphics[width=1\textwidth]{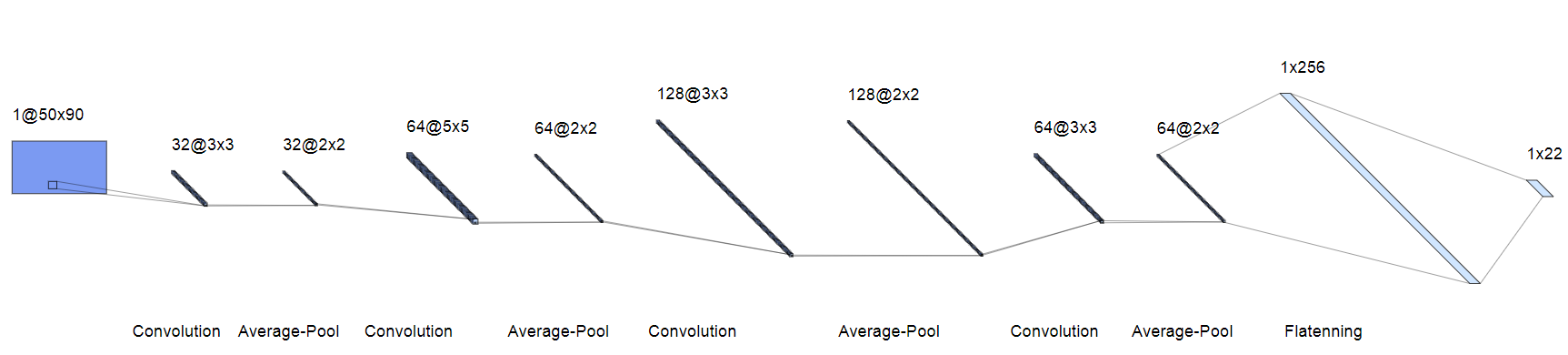}
	\caption{Implemented CNN Architecture}
	\label{fig:implemented_ML_model}
\end{figure*}

The used CNN is summarized in Fig.~\ref{fig:implemented_ML_model} where an overview of the implemented model is given as a block diagram. This model was inspired by the work in \cite{Nishio2018}. In \cite{Nishio2018} the inputs are also only the depth images of the environment and the output is a single integer (received power) that depends on the wireless channel.
 
We have observed that not using Batch Normalization layer which is used in the models in \cite{Nishio2018} decreased the training time without changing the performance for our data. In order to decrease the total number of parameters to prevent making the model more complex than needed and to further decrease the training time some more pooling layers are added. At the end the effect of kernel sizes and the number of filters in the convolutional layers on the performance were examined and the best performance is obtained with the given CNN structure in Fig.~\ref{fig:implemented_ML_model}.

The layers in Fig.~\ref{fig:implemented_ML_model} should be briefly explained before stating the hyperparameters in the training phase of the ML model. The first 2D Convolutional layer has 32 number of filters with kernel size of 3x3, where the rest of the Convolutional layers can be inferred from these sizes and the numbers for the other layers in Fig.~\ref{fig:implemented_ML_model}. All the 2D Average pooling layers has the same pooling size of 2x2. Note that Max pooling layers were also investigated and slightly better performance was observed when Average pooling layers were employed. ReLU is used after each convolution layers and after the first fully connected (dense) layer. The first dense layer has 256 neurons where it was seen that without this layer or increasing/decreasing the neuron number resulted in slightly worse performance. The last dense layer which is the output layer has 22 neurons and the selection of this neuron number is related to number of desired taps in the channel estimation. 

The hyperparameters of the given CNN model were selected as in \cite{Nishio2018} then tuned for better performance. The proposed CNN model's optimizer is selected as Nadam optimizer with an initial learning rate of 0.0001 and the learning rate is dropped to its 0.996 (decay=0.004) after every epoch which is the default value for Nadam optimizer. Total number of epochs is selected as 200. However, the model is selected according to the best performance on the validation set such that the ML model weights after a specific epoch that give best validation set performance are saved and used for evaluation. The loss metric for evaluating the validation set performance was selected as mean squared error. 

Average computation time of one channel estimation prediction is approximately 0.9 ms with the proposed CNN model. However, the time is observed in Jupyter Notebook (Python 3.6) operating with GeForce GTX 850 GPU. The trained model is also loaded to MATLAB operating with Intel Core i7-4700HQ CPU-2.40 GHz to predict the estimations where the average computation time is increased to approximately 9.8 ms.\footnote{Averages are calculated over 1000 channel estimation predictions.}

It should be pointed out that there exists also a pre-processing on the outputs of ML model before training which is the normalization of the channel impulse response CIR provided to output layer. The normalization is performed by dividing the CIR values by the maximum absolute valued CIR in the training set for each set combination. This absolute value is stored to revert the normalization during the evaluation of the comparison metrics on the test set since the algorithm provides normalized channel estimations.

%% file: chan_chan.tex
\section{Comparison}
\label{sec:comparison}

In this work wireless CIR is represented and estimated as FIR filter. The LS estimation forms the basis of all techniques. Although there exist more complex yet better estimations that take noise into account (such as minimum mean square error (MMSE) estimation), investigation of other techniques are left as future work to keep the proof of image based channel estimation simple. 

For the decoding of the packets, equalization is applied the same way for all estimation techniques. In other words, the only difference between compared techniques stems from the estimation part. Following, the compared techniques to Image Based Channel Estimation (VVD), are explained which are Data Based Channel Estimation and Kalman Filtering Based Channel Estimation.

The implemented and compared channel estimation techniques are:
\begin{itemize}
	\item Standard Decoding (No channel estimation),
	\item Ground Truth (Perfect channel estimation),
	\item Preamble Based,
	\item Preamble Based-Genie,
	\item 100ms Previous Estimation,
	\item 500ms Previous Estimation,
	\item Kalman AR(1),
	\item Kalman AR(5),
	\item Kalman AR(20),
	\item VVD-Current,
	\item VVD-33.3ms Future,
	\item VVD-100ms Future,
	\item Preamble-VVD Combined,
	\item Preamble-Kalman Combined.
\end{itemize} 

\subsection{No Estimation}

\textbf{Standard Decoding} indicates the standard of IEEE 802.15.4 where no channel estimation or equalization is utilized. Hence, in standard decoding only frequency offset correction and packet frame synchronization is performed.\footnote{Frequency offset correction and packet frame synchronization are performed in all other techniques as well.} In all other techniques ZF equalization is performed by using the estimated channel with the respective technique. Therefore, the only difference between the compared techniques stems from the difference in the estimation. 

\subsection{Data Based Channel Estimation}

Data based channel estimation corresponds to the training based channel estimation. It consists of 3 main parts, namely Perfect Channel Estimation, Preamble based Channel Estimation, Previous Channel Estimation in our work.

Previously, LS estimation is defined using the matrix $X^k$ where the parameter $M$ is indicating the known number of samples in the received signal.

In \textbf{Ground Truth}, the perfect channel estimation is utilized. The Perfect Channel Estimation is obtained by performing LS estimation to the whole received signal.\footnote{Number $M$ that is used in Equation \ref{eq:LS_estimate} becomes the total number of samples in a packet} This is practically impossible since it would mean that the receiver already knows the complete signal before decoding. Although this is not practical for communication between sensors, it is implemented to provide as a baseline for all other techniques.   

\begin{figure}[t!]
	\centering
	\includegraphics[width=0.42\textwidth]{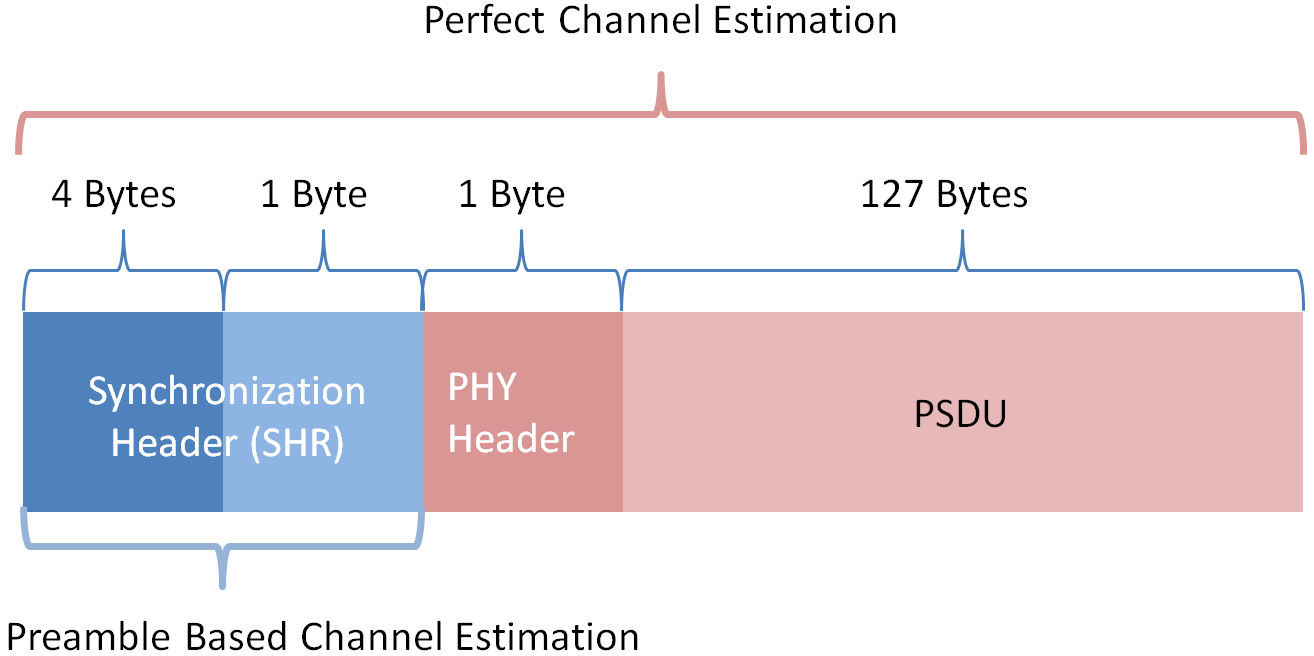}
	\caption{The considered reference parts of each received signal for Perfect and Preamble Based Channel Estimations}
	\label{fig:Preamble_Perfect_Est}
\end{figure}

\textbf{Preamble based} channel estimation is the practical version of the perfect channel estimation and exploits the already known synchronization field of the IEEE 802.15.4 payload as illustrated in Fig.~\ref{fig:Preamble_Perfect_Est}. However, this provides an estimation only if the preamble is decoded by the receiver. 

In \textbf{Preamble Based-Genie} technique, preamble is assumed to be always decoded. Thus, Preamble Based-Genie technique provides channel estimation from preamble part of the signal for every signal.

The \textbf{Previous Estimation} makes use of the perfect channel estimation from a previous signal. Hence, previous channel estimation is essentially a blind channel estimation for the signal of interest. There are two versions of previous channel estimation which are $100$ ms and $500$ ms previous channel estimations to reflect the effect of aging channel estimation. Both techniques assume that there exist always a clean packet reception within the defined interval.

\subsection{Kalman Filtering Based Channel Estimation}

The implemented Kalman filtering based channel estimation is a semi-blind channel estimation technique. In this work Kalman filtering based channel estimation is used as a blind channel estimation only to predict the future CIR. The implementation of the Kalman Filter is explained in Appendix. 

In order to have a fair comparison between VVD and Kalman filtering based channel estimation, the correlation coefficients in the Kalman filtering based channel estim	q	q1ation are calculated by using the channel estimates for each data set. After estimating the correlation coefficients of the wireless channel from training sets, Yule-Walker equations are used to calculate the coefficients of Auto-regressive (AR) process. Although it is known that Kalman filter converges to steady state quickly, first 200 packets in each set was not considered in the comparison results in order not to hinder the performance of Kalman filtering based channel estimation.

The variants of Kalman filtering based channel estimation are based on the order of AR process. Three different versions are implemented which are AR(1), AR(5), and AR(20). Kalman AR($p$) indicates a Kalman filtering based channel estimation with AR process of order $p$. The AR process order selection represents a trade of such that higher $p$ selection results in more accurate yet higher computational complexity \cite{Jamoos2008}.

On the other hand, variants of the VVD estimations are based on the prediction time. In detail, VVD has three variants where one version predicts the current wireless channel estimation, second version predicts the 33.3 ms into the future channel estimation, and the 100ms into the future channel estimation from the current camera images. All variants were trained separately but with the same algorithm. The only difference between the versions of VVD estimations is the output layer. In output layer the respective channel estimations for the targets were provided. In other words, providing input as the same image, the current channel estimation or 33.3 ms future channel estimation or 100 ms future channel estimation were given as outputs to the ML algorithm regarding the variant.

\subsection{Combined Channel Estimation}

\begin{figure}[t!]
  	\centering
  	\includegraphics[width=0.36\textwidth]{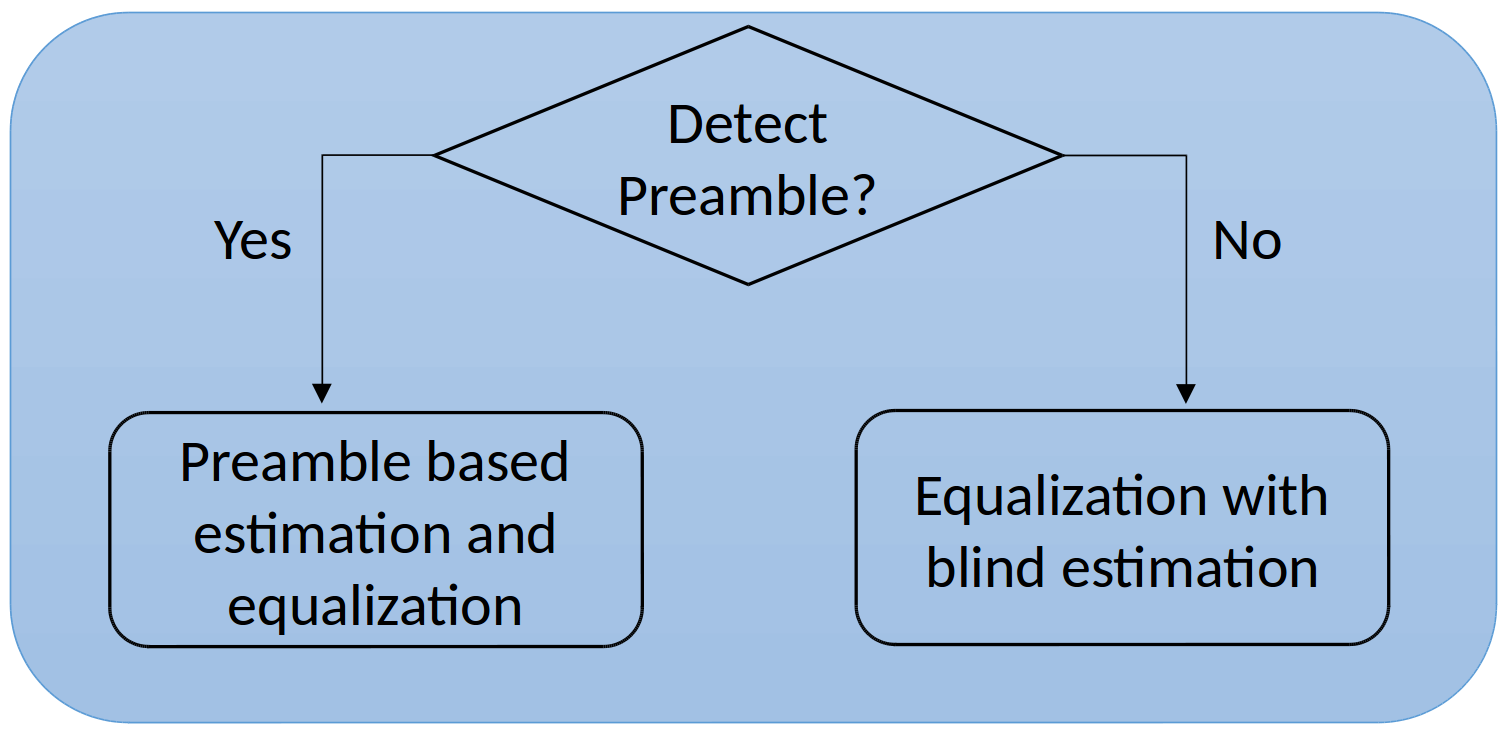}
  	\caption{Flow diagram for Combined Estimation}
  	\label{fig:comb_tech}
  \end{figure}

To use the advantage of both worlds, we have used the strength from equalization of blind estimation whenever there is a failure in the detection of the preamble. The flow diagram for the combination of blind estimations and preamble based estimation is illustrated in Fig.~\ref{fig:comb_tech}. These techniques are called combined estimation techniques. \textbf{Preamble-VVD} based combined technique is the combination of Preamble based and VVD-Current techniques. \textbf{Preamble-Kalman} Combined technique is the combination of Preamble based and Kalman AR(20).

\subsection{Metrics}

In this section, the metrics that are used for the comparison of different channel estimation techniques are introduced. Several metrics were selected in order to compare the performances of implemented channel estimation techniques in different aspects. The results obtained with these metrics are provided in Section 5.

In all of the compared channel estimation techniques ZF equalization which is explained in Section 2.2 is implemented. ZF equalization is applied whenever an estimation is available since the ZF approach requires the channel estimation. For blind channel estimation techniques there is no obstruction to apply ZF since whenever the receiver receives a signal the estimation is already known and does not depend on the received signal.

With blind estimation techniques\footnote{The techniques that provide the estimations regardless of the information from the received signal such as Kalman filtering based channel estimation, VVD, and previous channel estimation techniques.} the equalization is performed on each received signal. However, the preamble based channel estimation relies on the detection of preamble such that if the preamble is not detected practically no estimation could be obtained. If the preamble is not detected the signal is assumed to be erroneous.

\subsubsection{Packet Error Rate}

Packet error rate (PER), also known as packet loss ratio if no retransmission method is implemented, is the most important metric when observing the reliability of any communication system. Hence, this metric is employed to observe the reliability of the implemented channel estimations. It is calculated by counting the erroneous packets and the obtained number is divided by the total transmitted packets in the observed period.

\subsubsection{Chip Error Rate}

In IEEE 802.15.4 standard packets are transmitted in chips where each groups of 4 bits are mapped to sequences of 32 chips \cite{IEEE_2003}. The chip error is more granular compared to PER as each packet is composed of $8128$ chips. It also reflects the effect of different burst error patterns. Investigating chip error rate (CER) is proposed as the second metric. CER is calculated by the ratio of the number of erroneous chips in an equalized signal to the total number of chips in the signal.
 
\subsubsection{Mean Squared Error}

Mean Squared Error (MSE) is a metric to observe how well the implemented channel estimation techniques matches with the ground truth. Since the basis for all the compared estimation techniques is the implemented perfect channel estimation, the performance bound for any implemented channel estimation technique in this work is the perfect channel estimation, namely ground truth. The MSE calculation for one test set can be written as follows:
\begin{equation}
MSE = \frac{\sum_{k=1}^{z}\sum_{l=1}^{n}(\boldsymbol{h}_{l}^{k}-\boldsymbol{\hat{h}}_l^{k})^2}{z.n},
\label{eq:MSE}
\end{equation}
where $z$ denotes the number of transmissions in the considered test set, $n$ denotes the number of estimation taps, $\boldsymbol{h}_{l}^{k}$ corresponds to the perfect channel estimation at the tap number $l$ for the $k^{th}$ transmitted signal, and $\boldsymbol{\hat{h}}_l^{k}$ is used for the compared estimation.

\begin{figure}[t!]
	\begin{subfigure}{0.23\textwidth}
		\includegraphics[width=1.12\textwidth]{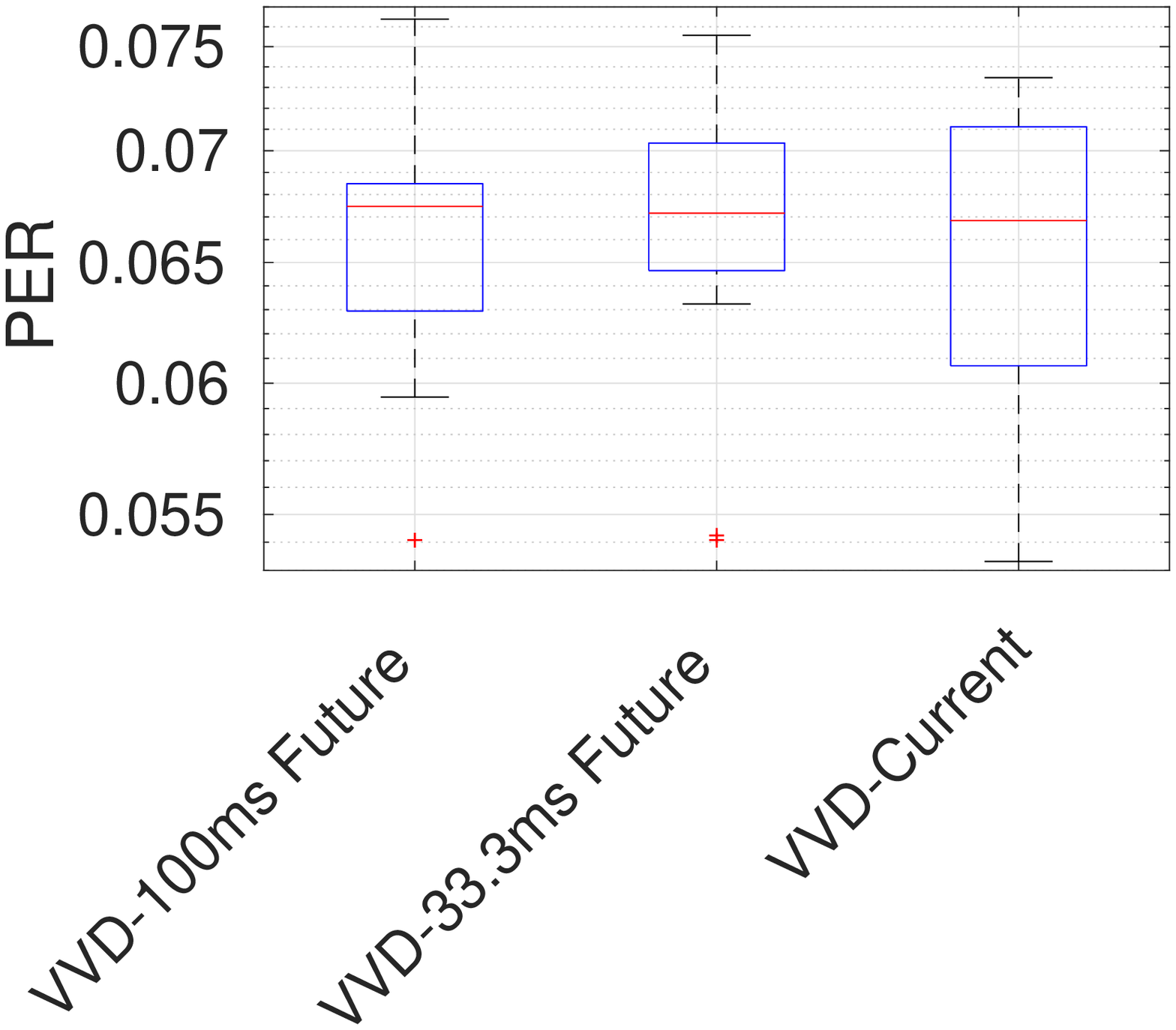}
		\centering
		\caption{VVD estimation}
		\label{fig:ML_PER}
	\end{subfigure}
	\hspace{-0.045cm}
	\begin{subfigure}{0.23\textwidth}
		\includegraphics[width=1.12\textwidth]{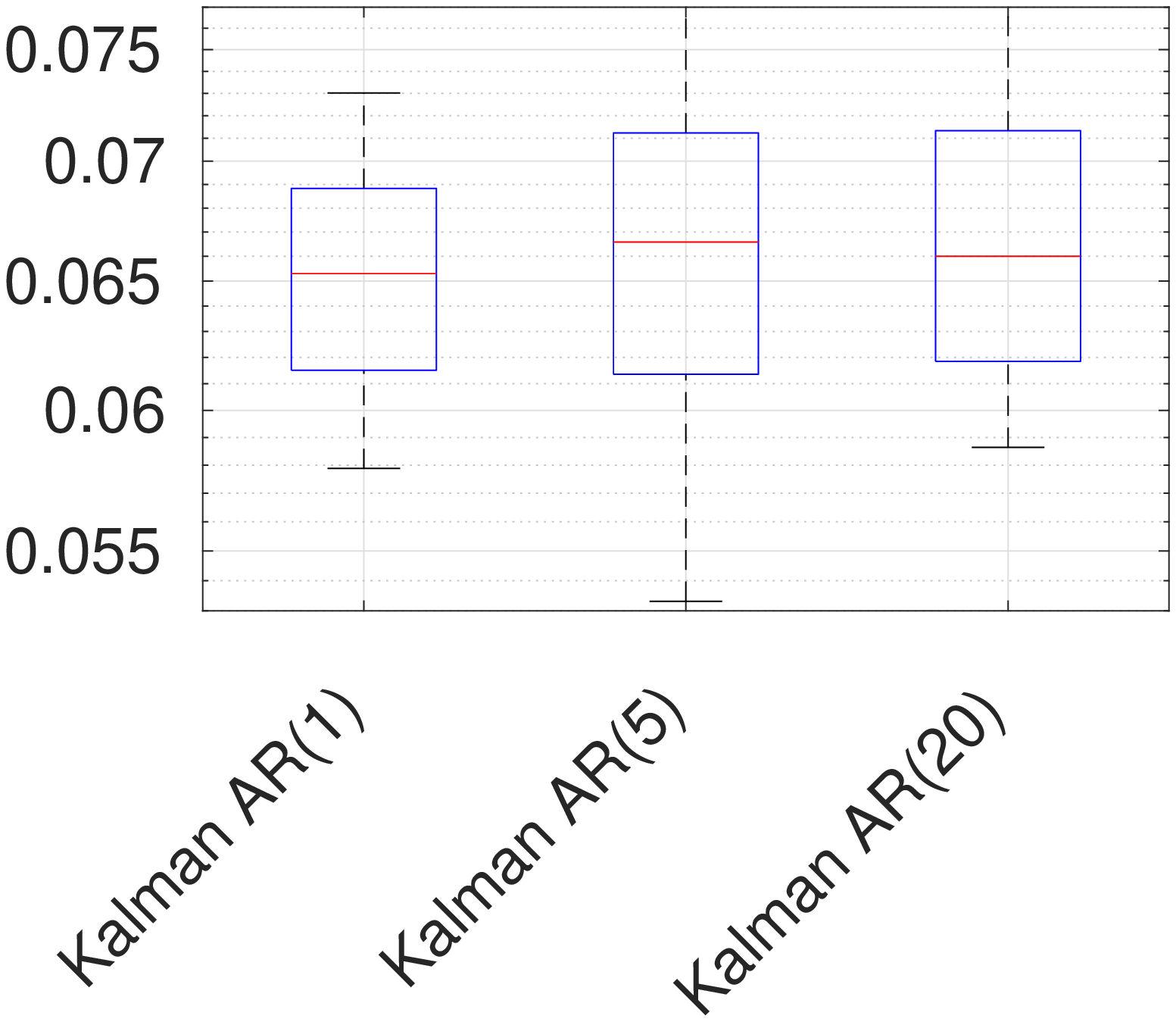}
		\centering
		\caption{Kalman estimation}
		\label{fig:Kalman_PER}
	\end{subfigure}
	\caption{PER comparison for variants of VVD and Kalman}
	\label{fig:Kalman_ML_PER}
\end{figure}

%% file: eval.tex
\section{Evaluations}
\label{sec:eval}

In this section the implemented channel estimation techniques are compared over test sets. The comparison results of different techniques are reported for the same test sets for each set combination. The set combination implies different partitioning of training, validation and test sets. Each combination has different test set in order to provide comparison results over each measurement take and have cross validation in ML results. Since the measurement was conducted in 15 takes, there exist 15 different set combinations. Evaluation with different set combinations is to increase validity of the results. The combinations that are used in evaluations are summarized in the Appendix.

As a final remark all the results for each comparison metric are provided over the mean of each test set for each set combination. In other words, the mean for each metric is calculated for each set combination which provides overall `15' mean values. Then, these mean values are put on box plots to visualize the distribution of the mean results of each set combination for each metric.  Furthermore, the variants of VVD algorithm based estimations are abbreviated as `VVD-Current', `VVD-33.3ms Future' and `VVD-100ms Future'.

\subsection{Packet Error Rate}

\begin{figure}[t!]
	\centering
	\includegraphics[width=0.52\textwidth]{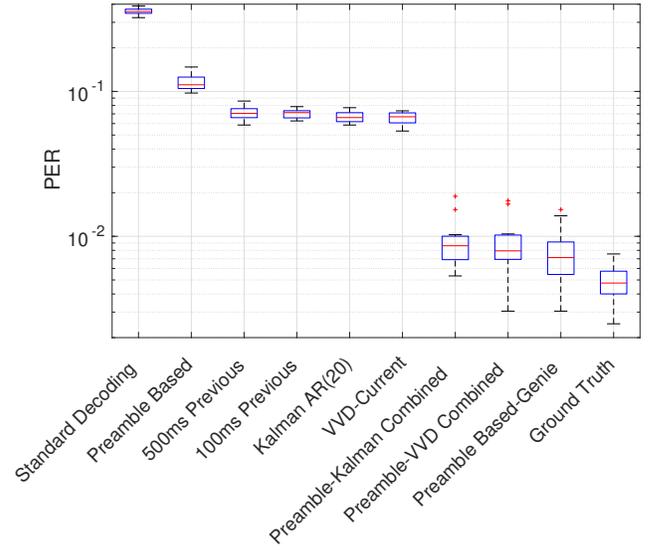}
	\caption{PER of all estimation techniques}
	\label{fig:PER}
\end{figure}

Initially we compare the variants of Kalman and VVD algorithms among themselves to select the best. In Fig.~\ref{fig:ML_PER} and \ref{fig:Kalman_PER} Packet Error Rate versus variants of each channel estimation algorithm is displayed. In Fig.~\ref{fig:Kalman_ML_PER}, it can be observed that all versions of Kalman estimation performed similarly. The result suggests that the wireless channel behaves almost memoryless. We believe this is due to the filter not differentiating LoS and NLoS scenarios. In Fig.~\ref{fig:ML_PER}, we observe that for the VVD algorithm inputting the recent image provides better channel estimation as expected. To avoid visual clutter in the overall comparison, VVD-Current and Kalman AR(20) are selected. %

\par The overall performance of each algorithm in terms of PER is illustrated in Fig.~\ref{fig:PER} as a box plot. We see that not utilizing equalization gives the worst performance while the ground truth scenario provides the best results. The Preamble Based estimation performs almost similar to blind estimation techniques, however, failure in detection of preamble holds it back. This effect is investigated in \cite{barac2014scrutinizing} where it is shown that up to 50\% of the packets in a measurement set are lost in the air due to preamble detection failure.

\par The PER performance for blind techniques are similar with slightly better performance for VVD. It should be noted that, all the blind estimation techniques requires a reception of a recent message except the VVD approach. Thus, they are usable for frequent periodic messaging but will face problems for sporadic messaging.
 
\par Finally, in the combined techniques where blind estimation is used to help decoding if the preamble detection is failed, practicality is the main consideration. If preamble is detected the preamble estimation is used, if not detected then the signal is equalized by using the blind estimations obtained with Kalman or VVD approaches. The combined techniques achieves almost two orders of magnitude decrease in packet error rate with respect to preamble based estimation. Both Kalman and VVD approaches behave similarly where VVD is better on average.

\subsection{Chip Error Rate}

\begin{figure}[t!]
	\centering
	\includegraphics[width=0.52\textwidth]{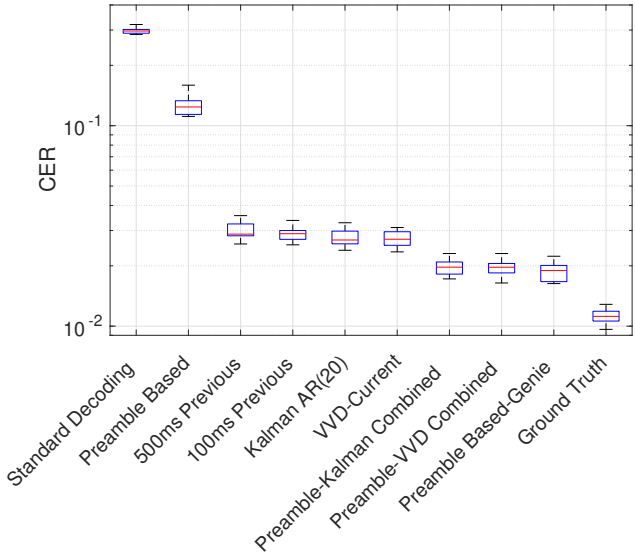}
	\caption{CER of all estimation techniques}
	\label{fig:CER}
\end{figure}

\begin{figure}[t!]
	\centering
	\includegraphics[width=0.52\textwidth]{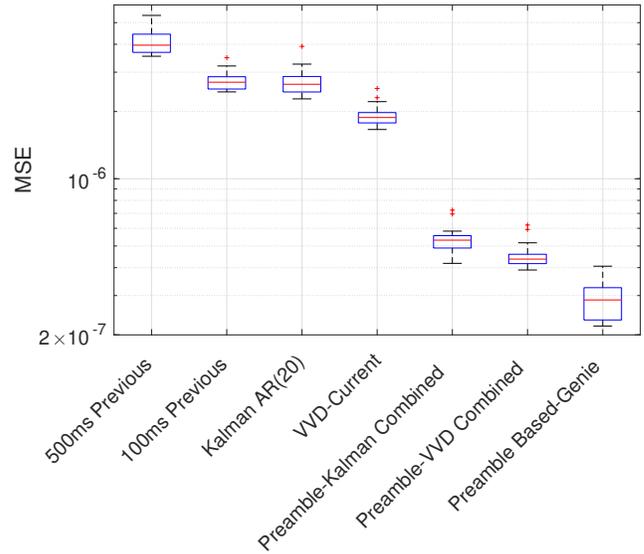}
	\caption{MSE of all estimation techniques}
	\label{fig:MSE}
\end{figure}

The CER for each estimation is summarized in Fig.~\ref{fig:CER}. Although the results displayed in Fig.~\ref{fig:CER} are observed to be in accordance with the results provided in Fig.~\ref{fig:PER} in general, the relative performance varies. The difference stems from having a Direct Spread Spectrum Sequence system where the bits are spread with pseudo noise PN sequences. The performance depends on the burst chip error pattern and the error correction probability of chip sequence to bit sequence mapping after each equalization technique. From the mapping, 32 chips are correlated with all of the 16 available PN sequences and the 4 bits of the most correlated PN sequence is selected. Note that the CER is calculated over 1016 bits in all of the measurements for all transmissions. Finally, the interesting gap between blind estimations and combined techniques have vanished. Although the gap is relatively small in CER, this small gap is almost an order of magnitude for PER. Hence, we project that there is an interesting CER threshold between $2\times10^{-2}$ and $3\times10^{-2}$ values that enables the receiver to map the chips to bits in a reliable fashion. 

\begin{figure}[t!]
	\centering
	\includegraphics[width=0.49\textwidth]{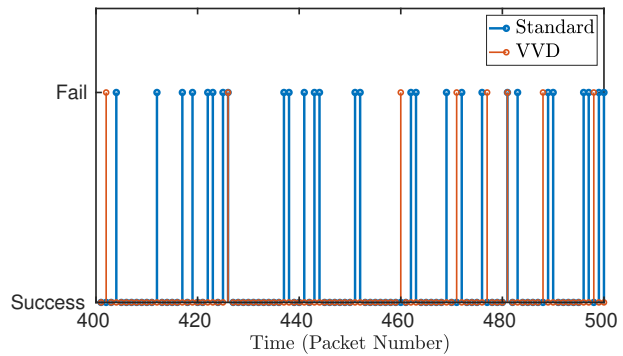}
	\caption{Time versus decoding performance}
	\label{fig:time_vs_decode}
\end{figure}

\subsection{Mean Squared Error}

MSE metric displays the channel estimation performances in higher granularity. From Fig.~\ref{fig:MSE}, it can be deduced that VVD approaches the baseline estimation approximately 1.4 times better than any other blind estimation technique. Furthermore, it can be observed that Preamble-VVD Combined technique is the best performing practical estimation technique similar to PER and CER results. Preamble based estimation is omitted in this plot since in case of preamble detection failure there exist no estimation to calculate MSE.

\subsection{Performance Insights}

In Fig.~\ref{fig:time_vs_decode} the packet failure versus time in the x-axis is illustrated. We investigate decoding of 100 packets. We observe a bursty behavior of packet errors which correlates with the human movement and blockage of line of sight. The VVD have errors transitioning to or from burst error regions such that it has hard time detecting if the movement will block the line-of-sight or not, at the edge cases. This provides insights into the behavior of VVD, such that better detection of a LoS and NLoS scenario can improve its performance.

\begin{figure}[t!]
	\begin{subfigure}[t]{0.985\linewidth}
		\includegraphics[width=\textwidth]{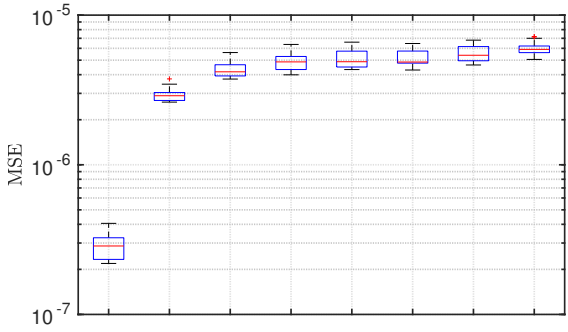}
		\centering
		\caption{Preamble Genie}
		\label{fig:gt_mse}
	\end{subfigure}
	\\
	\begin{subfigure}[t]{0.985\linewidth}
		\includegraphics[width=\textwidth]{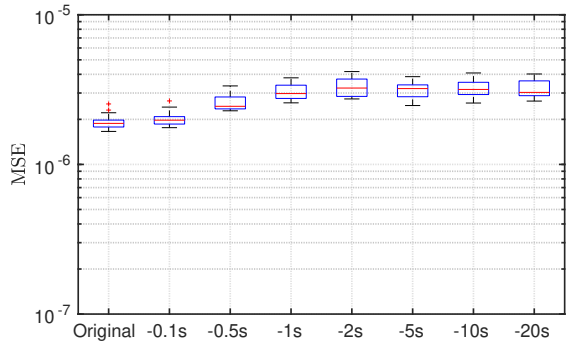}
		\centering
		\caption{VVD}
		\label{fig:ml_mse}
	\end{subfigure}
	\caption{Aging effect on mean squared error}
	\label{fig:compare_ageing_2}
\end{figure}

\subsection{Aging Insights}

\begin{figure}[t!]
	\begin{subfigure}[t]{0.985\linewidth}
		\includegraphics[width=\textwidth]{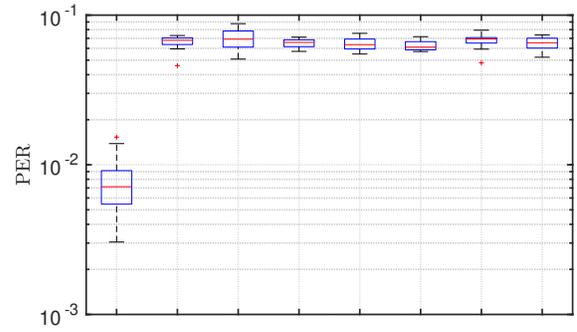}
		\centering
		\caption{Preamble Genie}
		\label{fig:gt_per}
	\end{subfigure}
	\\
	\begin{subfigure}[t]{0.985\linewidth}
		\includegraphics[width=\textwidth]{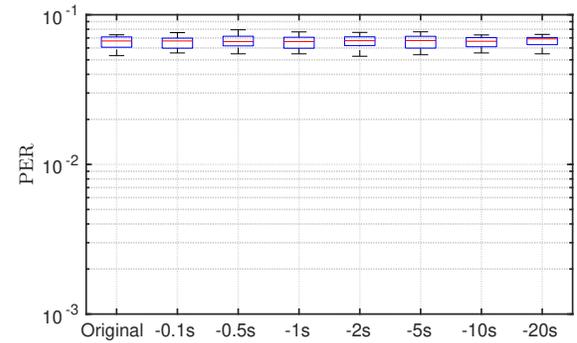}
		\centering
		\caption{VVD}
		\label{fig:ml_per}
	\end{subfigure}
	\caption{Aging effect on packet error rate}
	\label{fig:compare_ageing}
\end{figure}

In order to validate the instantaneous value of the information we have used an old channel estimation to either compare the difference with the recent channel estimation or to decode a recent packet. We mainly compare the VVD with the preamble genie estimation for this hypothesis.

In Fig.~\ref{fig:compare_ageing_2} we have plotted the mean squared error of two channel estimations while varying the age of the information from 0.1 second to 20 second. We see that in Fig.~\ref{fig:gt_mse} the error for the preamble genie increases from $2.9\times10^{-7}$ to $6\times10^{-6}$ exponentially with increasing age and stays at that value after age of 2 seconds. In Fig.~\ref{fig:ml_mse} we see that same trend exist for VVD with increasing from $1.9\times10^{-6}$ to $5\times10^{-6}$. These results validate the hypothesis for the effect of aging for channel estimation. 

In Fig.~\ref{fig:compare_ageing} we have plotted the effect of aging of packet error rate. Interestingly, we see observe a binary behavior for both estimation techniques. Especially, the effect of aging on PER for VVD 
is negligible as seen from Fig.~\ref{fig:ml_per}. For the preamble genie we observe a huge jump from $0.07$ to $0.2$ PER for 100 ms of aging . It is important here to consider that MSE is already at too high values to make a significant difference for PER such that the variations do not translate into a trend in terms of PER. This is due to the redundancy in the modulation technique as it is QPSK. We project that the reduction in MSE might be more meaningful for PER results for high-order modulation techniques. 

\subsection{Discussions}
\label{sec:discs}
\par All in all, these results partially confirm the main hypothesis of the work. The VVD algorithm behaves better than any time-series based estimation and currently used decoding. However, it does not match the ground truth information. The reason for this gap can be the size of the dataset or the CNN architecture. We believe that possibility of using images for channel estimation in factories and in-aircraft communication where the reliability of wireless information is crucial, is validated through this work. 

The proposed VVD algorithm is especially beneficial for keeping reliable channel estimations in use-cases with sporadic transmissions and long battery life such as safety-critical sensors. Time-series based estimations would require periodic pilot transmission to keep a reliable channel estimate and drain the battery or otherwise may fail due to lack of updates in channel estimations. Although beneficial in certain scenarios, VVD might have some limitations related to external interference, camera coverage, runtime, operating frequency and many more that should be carefully considered for system design.

VVD requires an interference-free environment to correctly learn the channel estimates if deployed in an online training and online inference fashion. However if VVD uses offline training in a interference-free environment and does online inference afterwards no interference-free environment is required during the operation.

In this work, VVD is investigated in an indoor environment for IEEE 802.15.4 standard. In principle, VVD should work in other wireless technologies, different bandwidths, varying transmission powers and with adaptive modulation and coding schemes. VVD could provide gain for adaptive modulation and coding schemes since they make use of varying channel states to maximize the information transfer. Typically, channel estimation performance varies with respect to mobility and interference. For this reason, encoder has to take into account that the channel estimation may fail due to these effects. As with VVD the decoder has a wireless channel estimate with a certain precision, the encoder can use this knowledge to deploy more aggressive modulation or coding schemes. Varying transmission power may increase the need for the dataset as the noise will be critical with decreasing power. Another aspect is the channel estimation technique used for the training phase e.g., LS estimation, the estimation we deploy, is not the best fit for low SNR region. Different signal bandwidth may affect the complexity of training data collection. However, after the channel estimations are extracted from the raw signals for training, no change in VVD is required. The effects of different modulation techniques on VVD are more complex. Intuitively, with higher modulation one would require more data to obtain more precise channel estimates as higher modulations are less tolerant to noise.

In the studied work, we assumed a block fading scenario. In other words, the channel varies slowly and assumed to be same for a certain duration called the coherence time. Depending on specific scenario this coherence time may vary and the real-time applicability of VVD depends on the particular coherence time. The coherence time in indoor environment depending on human speed can be assumed around 50 ms \cite{React}\footnote{Double the timing in the reference since our center frequency is half of \cite{React}.}. Considering that the VDD inference run-time is below 10 ms, even if CPU is used, VVD could be run in real-time. Of course this claim has to be validated with implementation. However, with faster objects the coherence time decreases and the inference runtime has to be improved. Since VVD can only provide estimation per camera-frame, depending on the coherence time of a specific environment frame rates may need to be increased and if the limits of cameras are reached more capable cameras may be needed.

We show that use of camera can improve the reliability by two orders of magnitude. However, one important gap to fill is how to conserve privacy of such a framework. One can argue that use of depth images is intractable to the human source but this may not always be the case. One solution is that the information is directly processed from pixels before they form an image \cite{yin2011information} and destroyed afterwards such that an image never forms. 

%% file: sota.tex
\section{Related Work}
\label{sec:sota}
In this section, related work on wireless communications is given for the considered approach of using ML to have a mapping from images to wireless channel information.

On the topic of machine learning usage for wireless reliable communication: 
In \cite{Ye2018} machine learning is used to extract channel estimation and for signal detection from received pilot symbols in an OFDM system. This scheme has advantages especially under severe channel conditions. This technique is meant for replacing signal processing techniques.

In \cite{Nishio2018} camera images are used to predict the received power in mmWave communications which could be assumed as the most relevant work to this approach. It is reported that in order to test their idea, in \cite{Nishio2018} two datasets were generated by simulation and by measurements. However, none of them are publicly made available. The suggested CNN based algorithm provides received power predictions using depth images. It should be noted that the implemented CNN model in our work is inspired by the work in \cite{Nishio2018}.

On the topic of wireless signal dataset: 
Specifically for IEEE 802.15.4, there doesn't exist a wireless waveform measurement dataset where the transmissions are from off the shelf sensors. The sole dataset \cite{Schmidt2018} is without wireless channel. For the collection of the data the transmitter and the receiver are connected with a coaxial cable. The authors in \cite{Schmidt2018} are not interested in the effects of wireless channel and the transmission is done from a vector signal generator instead of a dedicated IEEE 802.15.4 sensor. Moreover, our dataset is the first dataset that involves depth images from the communication environment together with the received wireless waveforms at the corresponding time instants.

%% file: Conc.tex
\section{Conclusion}
\label{sec:conc}

In this work, we have introduced Veni Vidi Dixi, VVD, a novel algorithm for blind wireless channel estimation using depth images of the communication environment. VVD is tested in an indoor scenario with a single human movement for a single transmitter receiver pair. In this scenario, VVD outperforms time-series based blind channel estimation techniques. As VVD does not require periodic transmission as in time-series based blind estimations, it avoids control message overhead for each added transmitter. Thus, it guarantees scaling with increasing number of transmitter.

This work is an initial proof of concept for using depth images for reliable wireless communication limited to the test scenario considered in this work. For generalization of the results, measurements have to be extended to multiple communication pairs and multiple mobile humans. However, it should be noted that the currently proposed VVD requires specific training for every different environment.

The authors believe that an important side information such as camera images should be used to improve communication reliability when available. This is especially intuitive for closed indoor scenarios where video surveillance is used such as factories or inside of an aircraft cabin. 

%% file: app.tex
\section*{Appendix}
\label{sec:Appen}

\subsection*{Kalman Filter Implementation}
\label{sec:KFI}
It is known that a fading wireless channel can be estimated with a Kalman filter as long as the channel is approximated by an Auto-regressive (AR) process \cite{Jamoos2008}. For a block fading channel with a $p^{th}$ order AR process, namely AR(p), $h_l^k$, the $l^{th}$ tap of the CIR at the received block $()^k$ can be demonstrated as \cite{Jamoos2008}\cite{Safaya2000}:

\begin{equation}
h_l^k = \sum_{i=1}^{p}\phi_ih_l^{k-i} + w_l^{k},
\label{eq:Kalman}
\end{equation}
where $\phi$ represents the coefficients of the AR model and $w_l^{k}$ denotes the white Gaussian noise.\footnote{Since a Wide Sense Stationary Uncorrelated Scattering (WSSUS) channel can be modeled as tapped delay line model which is our approach of estimating the channel, this WSSUS assumption indicates independent fading of the taps \cite[p.~112]{Molisch2011}. Thus, the taps of the Kalman filter is modeled independently.} The summation in Equation \ref{eq:Kalman} can be written in a matrix form as in,

\begin{equation}
\boldsymbol{h_l}^{k} = \boldsymbol{{\phi}_l}\boldsymbol{h_l}^{k-1} + \boldsymbol{w_l}^{k}.
\label{eq:Kalman_mat}
\end{equation}

$\boldsymbol{h_l^{k}}$ and $\boldsymbol{w_l^{k}}$ is a vector of length and $p$, $\boldsymbol{\phi_l}$ is a matrix with size $(p \times p)$. These can be summarized as:

\begin{gather*}
\hspace{-0.6cm}
\boldsymbol{h_l^{k}} = \begin{bmatrix}
h_l^{k} \\
h_l^{k-1} \\
h_l^{k-2} \\
... \\
h_l^{k-p+2} \\
h_l^{k-p+1}
\end{bmatrix},
\ 
\boldsymbol{w_l^{k}} = \begin{bmatrix}
w_l^{k} \\
0 \\
0 \\
... \\
0 \\
0
\end{bmatrix},
\boldsymbol{\phi_l} = \begin{bmatrix}
\phi_1 & \phi_2 & \phi_3 & ... & \phi_{p-1} & \phi_p \\
1 & 0 & 0 & ... & 0 & 0 \\
0 & 1 & 0 & ... & 0 & 0 \\
... & ... & ... & ... & ... & ... \\
0 & 0 & 0 & ... & 0 & 0 \\
0 & 0 & 0 & ... & 1 & 0 \\
\end{bmatrix}
\label{eq:Kalman_matrices}
\end{gather*}

The coefficients of AR model $\phi_1,\phi_2,\phi_3\ ...\ \phi_p$, represents a vector which is denoted as $\boldsymbol{\phi^p}$. In literature there exist several methods to estimate AR coefficients such as using the Jakes power spectrum by using the Jakes channel shaping filter \cite{Safaya2000}. However, Jakes approach requires the estimation of maximum Doppler frequency which is not guaranteed to be a trivial task \cite{Jamoos2008}. In another approach, which is implemented in our work, the AR model coefficients are calculated by utilizing the auto-correlation coefficients of the perfect channel estimates with Yule-Walker equations \cite{Jamoos2008}. Relation of autocorrelation coefficients of the perfect channel estimate with $\boldsymbol{\phi^p}$ for one tap can be represented in Eqs.~\ref{eq:R_phi}-\ref{eq:r_R2} by using the knowledge from Eq.~\ref{eq:Kalman} \cite{Safaya2000}:

\begin{equation}
R_l[\tau] = \sum_{i=1}^{p}\phi_iR_l[\tau-i] ,
\label{eq:R_phi}
\end{equation}
where $R_l$ corresponds to the autocorrelation of channel estimations, autocorrelation coefficient is given by:

\begin{equation}
r_l[\tau] = \frac{R_l[\tau]}{\sigma_{h_l}},  
\label{eq:r_R2}
\end{equation}

$\sigma_{h_l}$ the variance of the CIR for the tap number $l$. Overall these and the $\boldsymbol{\phi_l^p}$ can be represented in a matrix form as:

\begin{gather}
\boldsymbol{\phi_l^p} = R_l^{-1}\boldsymbol{r_l}.  
\label{eq:r_R}
\end{gather}
The vectors and matrices can be summarized as:

\begin{gather*}
\boldsymbol{\phi_l^p} = \begin{bmatrix}
\phi_1 \\
\phi_2 \\
... \\
\phi_p
\end{bmatrix},
\ 
\boldsymbol{r_l} = \begin{bmatrix}
r_l[1] \\
r_l[2] \\
... \\
r_l[p]
\end{bmatrix}
\label{eq:AR_matrices}
\end{gather*}
\begin{gather*}
R_l = \begin{bmatrix}
1 & r_l[1] & r_l[2] & r_l[3] & ...  & r_l[p-1] \\
r_l[1] & 1 & r_l[1] & r_l[2] & ...  & r_l[p-2] \\
... & ... & ... & ... & ... & ... \\
r_l[p-1] & ... & ... & ... & ... & 1 \\
\end{bmatrix}
\label{eq:AR_matrices}
\end{gather*}

The Kalman filter then can be written from Equations \ref{eq:Kalman}-\ref{eq:Kalman_mat}. The current channel estimate is given with:
\begin{equation}
\boldsymbol{\hat{h}_{l_{curr}}^{k}} = \boldsymbol{\hat{h}_l^{k}}+{K_l}^{k}(\boldsymbol{h_l^{k}}-\boldsymbol{\hat{h}_l^{k}}),
\label{eq:update1}
\end{equation}
where ${K_l}^{k}$ is the Kalman gain. The gain is calculated with
\begin{equation}
K_l^{k} = P_l^{k}(P_l^{k}+U)^{-1}, 
\end{equation}
where ${P}$ denotes the matrix for the predicted error covariance ${U}$ denotes the observation noise covariance. The update function for error covariance is
\begin{equation}
P_{l_{curr}}^{k} = (I - K_l^{k})P_{l}^{k},
\label{eq:update2}
\end{equation}
with ${I}$ corresponds to the identity matrix of size $pxp$.

Following, the predicted estimate is calculated with
\begin{equation}
\boldsymbol{\hat{h}_{l}^{k+1}} = \boldsymbol{\phi}_l\boldsymbol{\hat{h}_{l_{curr}}^{k}},
\label{eq:predict1}
\end{equation}
and the prediction of error covariance is updated with:
\begin{equation}
P_{l}^{k+1} = \boldsymbol{\phi_l}\boldsymbol{\hat{h}_{l_{curr}}^{k}}\boldsymbol{\phi_l}^H+Q,
\label{eq:predict2}
\end{equation}
where ${Q}$ denote the process noise covariance. Both ${U}$ and ${Q}$ assumed to be the same for all taps\footnote{We are updating the Kalman filter equations with the "perfect channel estimations", the observation noise is actually neglected by assigning small values. Since this is done for all taps ${U}$ is kept the same.}$^,$\footnote{The process noise covariance, also named as plant noise covariance, ${Q}$ is derived from previously defined $\boldsymbol{w_l}$. The mathematical representation is \(Q=E[\boldsymbol{w_l}\boldsymbol{w_l}^H]\)}. Since the deriving noise is same for all taps, ${Q}$ is kept same. The Equations \ref{eq:update1} and \ref{eq:update2} are named as the update steps of the Kalman filter, whereas the Equations \ref{eq:predict1} and \ref{eq:predict2} are called as the prediction steps of the Kalman filter.

\newpage

\subsection*{Set Combinations}
\label{sec:SC}

The table below summarizes how are the set combinations created from the measurements sets.  
\begin{table}[!htp]
	\footnotesize
	\centering
	\begin{minipage}{0.48\textwidth}
		\centering
		\begin{tabular}{|M{1.35cm}|M{2.65cm}|M{1.05cm}|M{0.35cm}|M{1.25cm}|N} 
			\hline
			\begin{minipage}{1.35cm} \vspace{0 cm} \centering \textbf{Combination} \\ \vspace{0 cm} \textbf{Number}\end{minipage} & \begin{minipage}{2.65cm} \vspace{0 cm} \centering \textbf{Training} \textbf{Sets}\end{minipage} & \begin{minipage}{1.05cm} \vspace{0 cm} \centering \textbf{Validation} \\ \vspace{0 cm} \textbf{Sets}\end{minipage} & \begin{minipage}{0.35cm} \vspace{0 cm} \centering \textbf{Test} \\ \vspace{0 cm} \textbf{Sets}\end{minipage} & \begin{minipage}{1.25cm} \vspace{0 cm} \centering \textbf{$\#$ of Packets} \\ \vspace{0 cm} \textbf{in Test Sets}\end{minipage}\\[7pt]
			\hline
			Combination 1 &  1,2,3,4,5,7,9,10,11,12,13,14,15 & 6 & 8 & \begin{minipage}{1.25cm} \vspace{0.01cm} \centering 1213\end{minipage}\\[1.2pt]
			\hline
			Combination 2 &  1,2,3,4,5,6,7,8,9,10,12,13,14 & 11 & 15 & \begin{minipage}{1.25cm} \vspace{0.01cm} \centering 919\end{minipage}\\[1.2pt]
			\hline
			Combination 3 &  1,2,3,4,5,6,7,8,10,11,12,13,15 & 14 & 9 & \begin{minipage}{1.25cm} \vspace{0.01cm} \centering 1223\end{minipage}\\[1.2pt]
			\hline
			Combination 4 &  1,3,4,6,7,8,9,10,11,12,13,14,15 & 5 & 2 & \begin{minipage}{1.25cm} \vspace{0.01cm} \centering 2319\end{minipage}\\[1.2pt]
			\hline
			Combination 5 &  1,2,3,5,6,7,8,9,10,11,13,14,15 & 12 & 4 & \begin{minipage}{1.25cm} \vspace{0.01cm} \centering 1478\end{minipage}\\[1.2pt]
			\hline
			Combination 6 &  2,3,4,5,6,7,8,9,11,12,13,14,15 & 10 & 1 & \begin{minipage}{1.25cm} \vspace{0.01cm} \centering 2822\end{minipage}\\[1.2pt]
			\hline
			Combination 7 &  1,2,3,4,5,7,8,10,11,12,13,14,15 & 9 & 6 & \begin{minipage}{1.25cm} \vspace{0.01cm} \centering 1460\end{minipage}\\[1.2pt]
			\hline
			Combination 8 &  1,2,4,5,6,7,8,9,10,11,12,14,15 & 13 & 3 & \begin{minipage}{1.25cm} \vspace{0.01cm} \centering 2609\end{minipage}\\[1.2pt]
			\hline
			Combination 9 &  1,2,3,4,6,7,9,10,11,12,13,14,15 & 8 & 5 & \begin{minipage}{1.25cm} \vspace{0.01cm} \centering 1397\end{minipage}\\[1.2pt]
			\hline
			Combination 10 &  1,2,3,5,6,8,9,10,11,12,13,14,15 & 4 & 7 & \begin{minipage}{1.25cm} \vspace{0.01cm} \centering 1513\end{minipage}\\[1.2pt]
			\hline
			Combination 11 &  1,2,4,5,6,7,8,9,11,12,13,14,15 & 3 & 10 & \begin{minipage}{1.25cm} \vspace{0.01cm} \centering 994\end{minipage}\\[1.2pt]
			\hline
			Combination 12 &  1,2,3,4,5,6,8,9,10,12,13,14,15 & 7 & 11 & \begin{minipage}{1.25cm} \vspace{0.01cm} \centering 1201\end{minipage}\\[1.2pt]
			\hline
			Combination 13 &  1,2,3,4,5,6,7,8,9,10,11,14,15 & 13 & 12 & \begin{minipage}{1.25cm} \vspace{0.01cm} \centering 1243\end{minipage}\\[1.2pt]
			\hline
			Combination 14 &  1,3,4,5,6,7,8,9,10,11,12,14,15 & 2 & 13 & \begin{minipage}{1.25cm} \vspace{0.01cm} \centering 1272\end{minipage}\\[1.2pt]
			\hline
			Combination 15 &  2,3,4,5,6,7,8,9,10,11,12,13,15 & 1 & 14 & \begin{minipage}{1.25cm} \vspace{0.01cm} \centering 1041\end{minipage}\\[1.2pt]
			\hline
		\end{tabular}
		\caption{All set combinations used in VVD comparison}
		\label{table:Set_Combinations}
	\end{minipage}
\end{table}